\documentclass{article}
\usepackage{arxiv}
\usepackage[utf8]{inputenc} % allow utf-8 input
\usepackage[T1]{fontenc}    % use 8-bit T1 fonts
\usepackage{hyperref}       % hyperlinks
\usepackage{url}            % simple URL typesetting
\usepackage{booktabs}       % professional-quality tables
\usepackage{amsfonts}       % blackboard math symbols
\usepackage{nicefrac}       % compact symbols for 1/2, etc.
\usepackage{microtype}      % microtypography
\usepackage{lipsum}		% Can be removed after putting your text content
\usepackage{natbib}
\usepackage{doi}
\usepackage{psfrag,color}
\usepackage[dvips]{graphicx}
\graphicspath{{./Figures/}} % declare the path(s) where your graphic files are
\DeclareGraphicsExtensions{.eps} % and their extensions to bypass within every instance of \includegraphics

\title{Admission Control for Inelastic Traffic on a Link Shared by Deadline-Driven Elastic Traffic}

\author{ {O.~Patrick Kreidl}\thanks{This manuscript is the unaltered 2018 submission to {\it IEEE Transactions on Networking}, which was rejected.} \\
	School of Engineering (with Courtesy Appointment in School of Computing)\\
	University of North Florida\\
	Jacksonville, FL 12224 \\
	\texttt{patrick.kreidl@unf.edu} \\
}

\date{}

\renewcommand{\headeright}{O.~Patrick Kreidl}
\renewcommand{\undertitle}{}
\renewcommand{\shorttitle}{Admission Control for Inelastic Traffic}

\nocite{*}

\newcommand{\FigRef}[1]{Fig.~\ref{fig:#1}}
\newcommand{\SecRef}[1]{Section~\ref{sec:#1}}
\newcommand{\sSecRef}[1]{Subsection~\ref{ssec:#1}}
\newcommand{\equRef}[1]{(\ref{equ:#1})}
\newcommand{\mat}[1]{\ensuremath{\mathbf{#1}}}

\begin{document}
\maketitle

\begin{abstract}
Consider a (logical) link between two distributed data centers with available bandwidth
designated for both deadline-driven elastic traffic, such as for scheduled
synchronization services, and profitable inelastic traffic, such as for real-time
streaming services. Admission control in this setting is cast as a stochastic shortest
 path problem, with state space derived from (discretization of) the elastic flow's
 size/deadline and action space corresponding to alternative subsets of admitted
 inelastic flows: the probabilistic model expresses uncertainty in both the link's available
 bandwidth and the inelastic flows' offered loads, while the objective function captures both
 congestion avoidance and the option to specify a desired minimum elastic rate. Its solution
 is shown to (i)~balance the accumulation of instantaneous inelastic reward with the
  risk of missing the elastic deadline and (ii)~exhibit a degree of robustness to
 link \& flow modeling errors that is tunable via choice of the desired minimum elastic rate. Also
 discussed are state augmentations that befit urgent or
 non-interruptible inelastic traffic.
\end{abstract}

\newpage
\section*{Ackowledgments}
\begin{quote}
``If at first you don't succeed, give up.'' \quad {\it ---Homer Simpson.}
\end{quote}

This manuscript is the unaltered 2018 submission to {\it IEEE Transactions on Networking}, which was rejected. The reviewers deemed the main model as overly-simplified and, in turn, the implications of the analysis of little practical relevance. Though the author recognized this manuscript's departure from the seminal convex programming formulation (e.g., \cite{KMT98:RatCo} of prior work, that the reviewers saw no value in this manuscript was a surprise. Unfortunately, the networking project supporting this work was too near its end to address the reviewers' concerns and, in turn, the manuscript remains unaltered. Due to the intellectual influence of this project on this manuscript, the project collaborators are gratefully acknowledged: they are Amin Aramoon, Jamie Floyd, Gregory Frazier and Jacob McGill of Apogee Research, LLC in Arlington, VA as well as Scott Alexander, Steven Beitzel, Gary Levin, Eric Van Den Berg and Gary Walther of Perspecta Labs 
(formerly Applied Communication Sciences) in Basking Ridge, NJ.

Today, it is \today ~and the author keeps no ambition to conventionally publish this manuscript, so is choosing to upload it to the \emph{arXiv} open-access online repository. The manuscript is essentially unreviewed other than those leading to the 2018 rejection decision. Readers are invited to communicate questions or errata to the author via email: {\tt patrick.kreidl@unf.edu}

\newpage
\setlength{\topmargin}{-0.5in}
\setlength{\headsep}{0.5in}
\pagestyle{myheadings}
\addtocounter{page}{-2}
\markright{Admission Control for Inelastic Traffic on a Link Shared by Deadline-Driven Elastic Traffic \hfill 
Kreidl--}

\newpage
\section{Introduction \label{sec:intro}}
Growth in demand for multimedia online services raises the
stakes for Internet control schemes that support configurable precedence
among multi-class traffic \cite{Wri07:AdCon,rfc11:n6077}. Perhaps the most prevalent
example in today's Internet concerns whether increasing volumes of inelastic RTP/UDP
traffic (e.g., VoIP, video), which is generically unresponsive to network congestion,
would interact friendly/fairly with the transport-layer congestion control mechanisms
built into elastic TCP traffic (e.g., email, web). While prior work \cite{BLT06:TCPFr,CCK08:ElIne}
predicted that such concerns may be slow to materialize, future
applications are expected to only push further beyond the assumptions of existing protocol
suites. Such anticipation is evident in ongoing work on multi-class rate
control \cite{LMS05:OptRC,CMW10:ElIne,VTH13:NetUM,LST16:ResAl,PhH17:CCSur}, for example, which extends the seminal work
treating only elastic traffic \cite{KMT98:RatCo,LaA02:UtBRC} by associating different
(albeit non-convex) utility functions to inelastic classes.

Because this growth in demand is global, multimedia service providers are increasingly
relying upon the use of geographically-distributed data centers. Datacenter networks
themselves are presenting new challenges to existing protocol suites, among them being
an increased emphasis on deadline-driven elastic services that arise in the context of
recurrent content synchronization tasks \cite{ZCB17:DeaDC}. Ongoing work in datacenter
networks is addressing some limitations of rate-controlled TCP in the presence of
deadlines \cite{ChP07:DeaNC,WBK11:DeaDC,LXC13:DeaDC,XLR15:DeaDC,HHC15:DeaDC,HZH17:DeaMP}. However, none
(to our knowledge) have explicitly
considered the possibility that bandwidth reserved for deadline-driven elastic
services may also be in contention with potentially profitable inelastic services.
Combined with the fact that rate allocation for deadline-driven traffic is often
conservative, so as to hedge against worse-than-expected future network conditions,
the opportunity to admit inelastic traffic during nominal or better-than-expected
network conditions ought not be dismissed.

\begin{figure}[!b]
\centering
\vspace{0.1in}
\includegraphics[width=0.45\columnwidth]{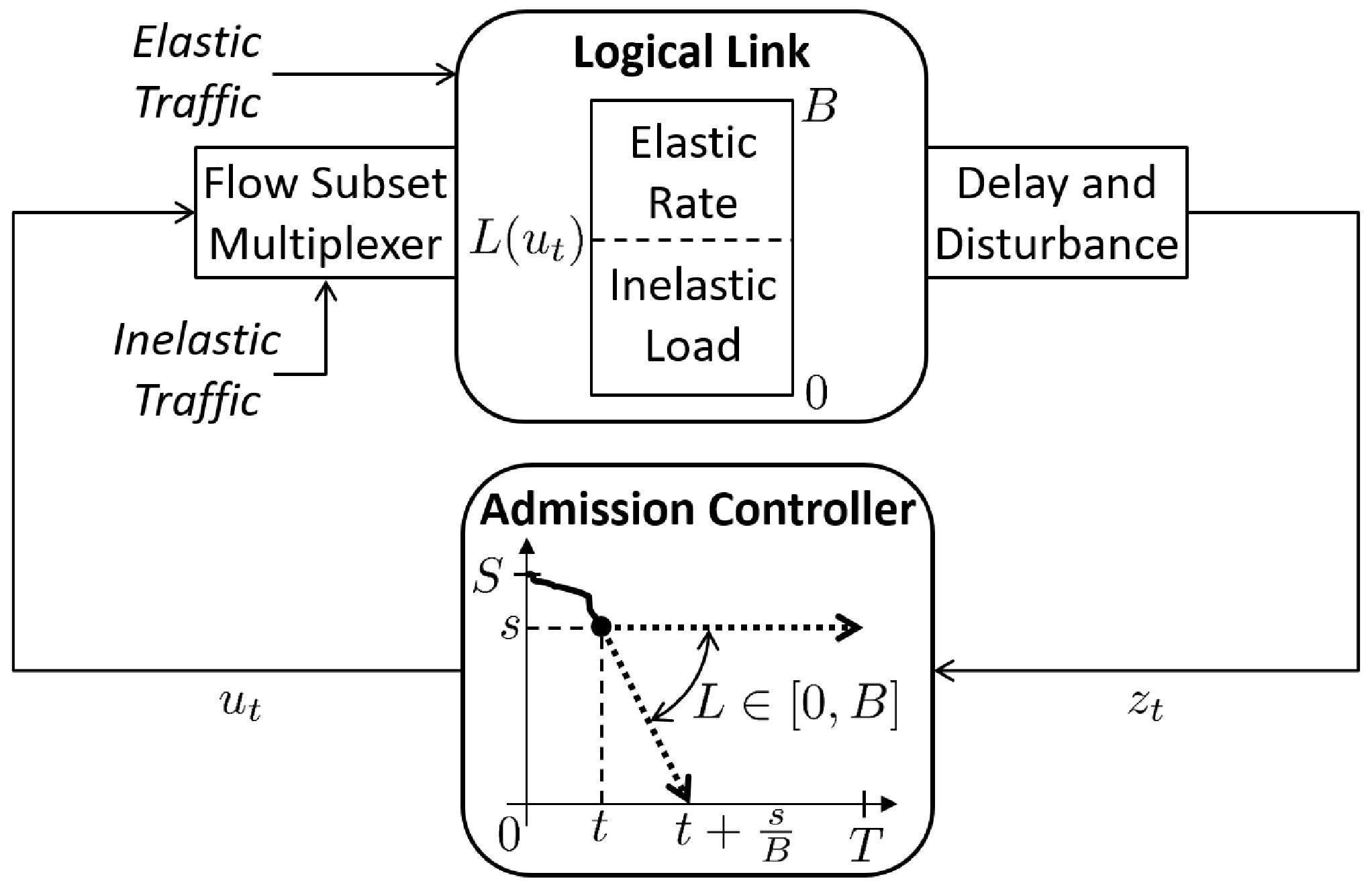}
\caption{\label{fig:MainSetup} Admission control for a link with available
bandwidth $B$ (in megabits per second) designated primarily for elastic traffic with size $S$ and deadline
$T$ (in megabits and seconds, respectively) in contention with profitable
inelastic traffic, each (discrete) control action $a$ imposing a mean offered load $L(a)$ (in megabits per second)
associated to the admitted subset of inelastic flows. The controller tracks the time-varying
status $z_t$, including the elastic flow's remaining size $s$ in $[0,S]$ and elapsed time $t$ in $[0,T]$,
to exert periodic control $u_t$ on the admitted subset of inelastic flows and, in turn, on the mean
elastic rate $B - L(u_t)$ into the subsequent time period. The control objective is to
maintain negligible risk of missing
the elastic deadline yet sustain the most profitable inelastic traffic.}
\end{figure}

This paper considers the setup illustrated in \FigRef{MainSetup}, assuming a logical link with available bandwidth %(in megabits per second)
designated for both a deadline-driven elastic flow %(with deadline and size in seconds and megabytes, respectively)
and a set of inelastic flows, each admissible subset associated with a mean offered load. %(in megabits per second); see \FigRef{MainSetup}.
\SecRef{ProbForm} casts admission control in this setting
as a stochastic shortest path problem \cite{Ber17:DPOC1}, a particularly well-understood
instance of Markov Decision Process (MDP) models that have been applied in numerous other
networking contexts e.g., \cite{Alt02:MDPCN}. Our probabilistic model quantifies the degree
to which a heavier total inelastic load, by reducing the effective bandwidth for the elastic flow or
even starving it out entirely under congestion, increases the risk of missing the deadline;
meanwhile, our objective function quantifies the balance between the accumulation
of instantaneous inelastic reward with the potential to forego the one-time reward of
completing the elastic flow within deadline. \SecRef{NomAnal} employs the formulation to analyze
specific scenarios with real-world link and traffic assumptions, illustrating the interplay of
key parameters in forming a tractable optimization problem as well as key insights behind the
achieved performance gains over uncontrolled (i.e., fixed-admission) policies. These analyses
also expose two limitations of
the core formulation, the first concerning robustness to modeling errors of link loads and the
second concerning absent constraints on inelastic services, both examined further
in \SecRef{AugmForm}. Specifically, \sSecRef{RobAnal} extends the formulation
to include a desired minimum elastic rate, providing a mechanism by which to tune
policies for varying degrees of robustness to mis-estimated link bandwidth or inelastic loads; \sSecRef{StateAug}
discusses state augmentations that befit urgent or non-interruptible
inelastic flows, albeit at the expense of increased model complexity. Conclusions and suggested future
research are provided in \SecRef{Conclu}.

\section{Stochastic Shortest Path Formulation \label{sec:ProbForm}}
This section describes the steps by which the sequential decision process depicted
in \FigRef{MainSetup} is cast as a standard stochastic shortest path problem \cite{Ber17:DPOC1}.
Such problems, assuming $n$ system states and $m$ control actions, are generically
modeled by two collections of (action-dependent) $n$-by-$n$ real-valued matrices,
\begin{equation}
\label{equ:SSPModel}
\mat{F}(a) \quad \mbox{and} \quad \mat{G}(a), \qquad a = 1,2,\ldots,m
\end{equation}
where the row-$i$, column-$j$ element of each such matrix  represents the
single-stage probability $F_{i,j}(a)$ and cost $G_{i,j}(a)$, respectively, of entering state $j$ conditioned on being
in state $i$ and applying action $a$. Note that the probabilistic
model is a set of stochastic matrices, each consisting of only non-negative entries that must
sum to unity across every row. Also assumed is the existence of a cost-free terminal
state, conventionally state $n$ and thus element $(n,n)$ is one in every stochastic
matrix and zero in every cost matrix. Control decisions are made in stages, as a function
of the evolving state, and costs accrue additively. Standard dynamic programming
algorithms (e.g., value iteration), minimizing the
expected total cost to reach termination by choice of (stationary) policy
$\mu:\{ 1,2,\ldots,n \} \rightarrow \{ 1,2,\ldots,m \}$, provably converge
provided there is at least one policy under which there exists a probable path
from every state to the terminal state.

\begin{figure}[b]
\centering
\vspace{0.1in}
\includegraphics[width=0.45\columnwidth]{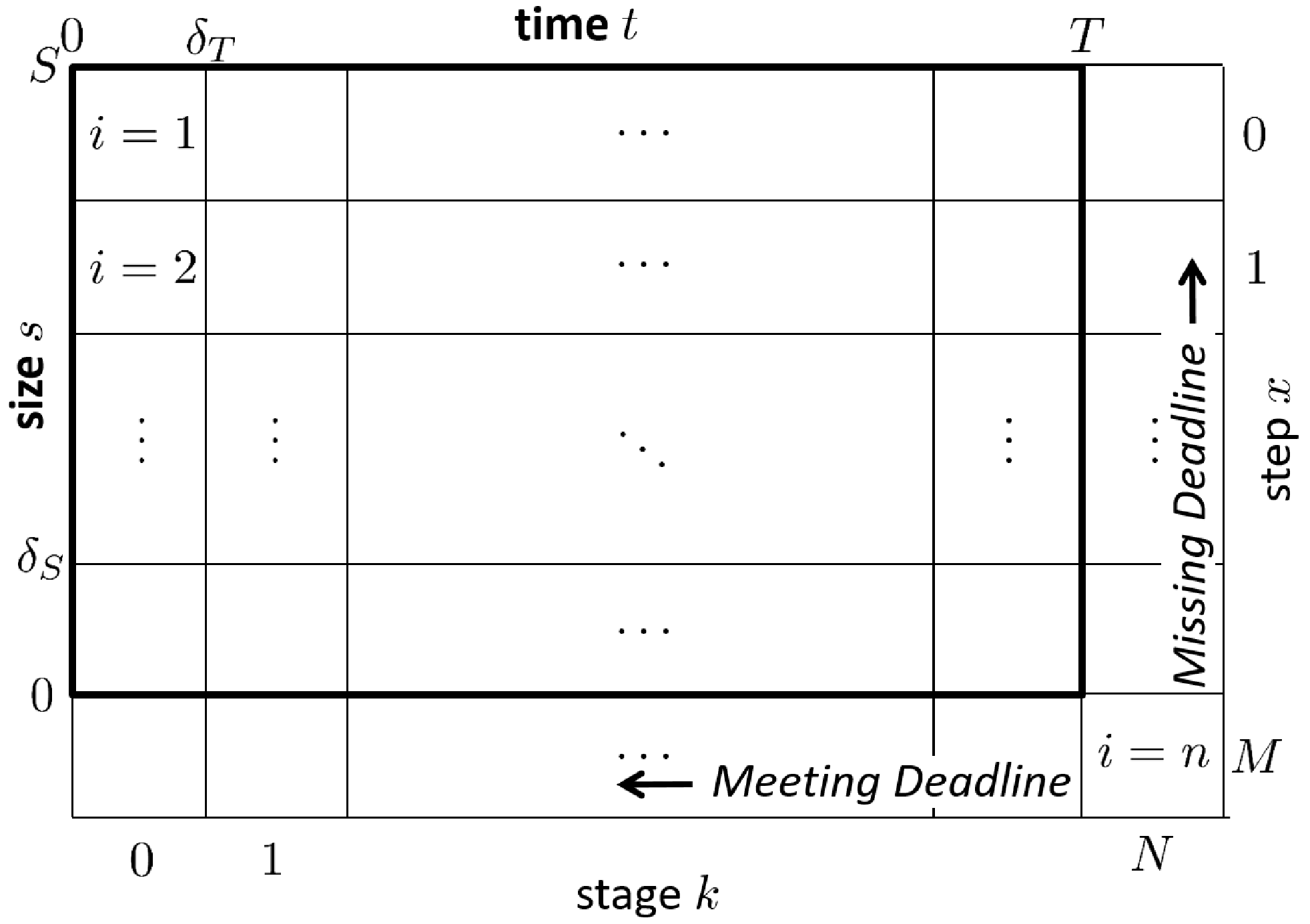}
\caption{\label{fig:StateSpace} Discretization of our admission controller's continuous-valued
state-space, viewing the elastic flow with size $S$ and deadline $T$ as
an $M$-step task to complete in at most $N \geq M$ stages. The resulting $n = (M+1)(N+1)$ cells
correspond to the finite set of states in our stochastic shortest path formulation, the index $i = 1, 2, \ldots, n$
assigned to cells taken column-wise from left-to-right and from top-to-bottom within each column.
Observe that the top-left cell (index 1) contains the initial status $(s,t) = (S,0)$, while the
bottom-right cell (index $n$) represents the cost-free terminal state and every other cell in the bottom row
or the right column represents meeting or missing, respectively, the deadline.}
\end{figure}

\subsection{From Scenario Parameters to Model Structure \label{ssec:TheStruc}}
The model structure of a stochastic shortest path problem refers to the (finite) cardinalities
$n$ and $m$ of the state space and control space, respectively. The setup of \FigRef{MainSetup}
introduces the following key scenario parameters: the
\begin{itemize}
\item link's available bandwidth $B$ (in megabits per second),
\item elastic flow's size $S$ and deadline $T$ (in megabits and seconds, respectively), and
\item inelastic traffic presented as a collection of candidate flows, each control action $a$
associated with admission of a distinct subset of those flows with mean offered load $L(a)$
(in megabits per second).
\end{itemize}
Also implied by \FigRef{MainSetup} is that the controller's notion of state includes a
continuous-valued quantity in two dimensions, the elastic flow's remaining size $s$ in $[0,S]$ and
elapsed time $t$ in $[0,T]$. However, the control space, in correspondence
with alternative subsets of %admitted
inelastic flows, is inherently discrete.

\subsubsection{Finite State Space}
Our discretization of the state space views the elastic flow as an $M$-step task to complete in at
most $N \geq M$ stages. Specifically, as \FigRef{StateSpace} illustrates, the size interval $[0,S]$ is divided
uniformly into $M+1$ steps indexed by $x = 0, 1, \ldots, M$ (assigning $\delta_S = S/M$ megabits/step), while
the time interval $[0,T]$ is
divided uniformly into $N+1$ stages indexed by $k = 0, 1, \ldots, N$ (assigning $\delta_T = T/N$ seconds/stage).
Along each dimension, we follow the convention that each index
is assigned the half-open interval of its associated continuous segment, and thus only a singleton set
remains for the final index. That is, each step $x < M$ corresponds to remaining size $s$ contained in
$\left( S-(x+1)\delta_S,S-x\delta_S \right]$ whereas $x = M$ corresponds to $s = 0$, or completing the
flow. Similarly, each stage $k < N$ corresponds to elpased time $t$ contained in
$\left[ k\delta_T,(k+1)\delta_T \right)$ whereas $k = N$ corresponds to $t = T$, or reaching the
deadline. The resulting grid of $n = (M+1)(N+1)$ cells is then
linearly indexed by $i  = 1 + x + (M+1)k$, resulting in corresponding orderings of
\begin{equation}
\label{equ:StateSpace}
\begin{array}{l}
\hspace*{-0.05in} (x,k) = (0,0), (1,0), \ldots, (M,0), (0,1),(1,1), \ldots, (M,N) \\
\hspace*{-0.05in} \qquad\:\:\,\, \updownarrow \\
\hspace*{-0.05in} \quad i \quad = \quad 1 \:\:\: , \quad 2 \:\:\:, \ldots, M+1, M+2, M+3, \ldots, n \:\:.
\end{array}
\end{equation}
It is worth noting that modeling the system dynamics approximately over the discrete state space raises an important
consideration in practice: for any particular scenario, discretizing
coarsely limits the controller's agility (and likely the achievable performance gains over fixed-admission
policies) but discretizing finely raises the problem's complexity.

\subsubsection{Finite Control Space}
In the size-$m$ control space, one action per admissible subset of inelastic flows, we
assume (i)~the empty set is always among the alternatives and (ii)~inelastic loads that
exceed the link's available bandwidth are never intensionally invoked.\footnote{Unintentional
admission of inelastic load in excess of available bandwidth will
be introduced in \SecRef{AugmForm}, when the impact of link/traffic modeling errors as well as possibly
detrimental congestion is examined.} Without
loss-of-generality, the actions can be indexed such that
$$
0 = L(1) < L(2) \leq \cdots \leq L(m) \leq B
$$
so, letting $R(a) = B - L(a)$, the associated action-dependent mean elastic rates are
\begin{equation}
\label{equ:ActionSpace}
B = R(1) > R(2) \geq \cdots \geq R(m) \geq 0.
\end{equation}
It is worth noting that choosing $m$ to index the full power set over all inelastic flows
is likely too large, so scenario-dependent schemes to construct more manageable control
spaces becomes another important consideration in practice.

\begin{figure}[b]
\centering
\vspace{0.1in}
\includegraphics[width=0.5\columnwidth]{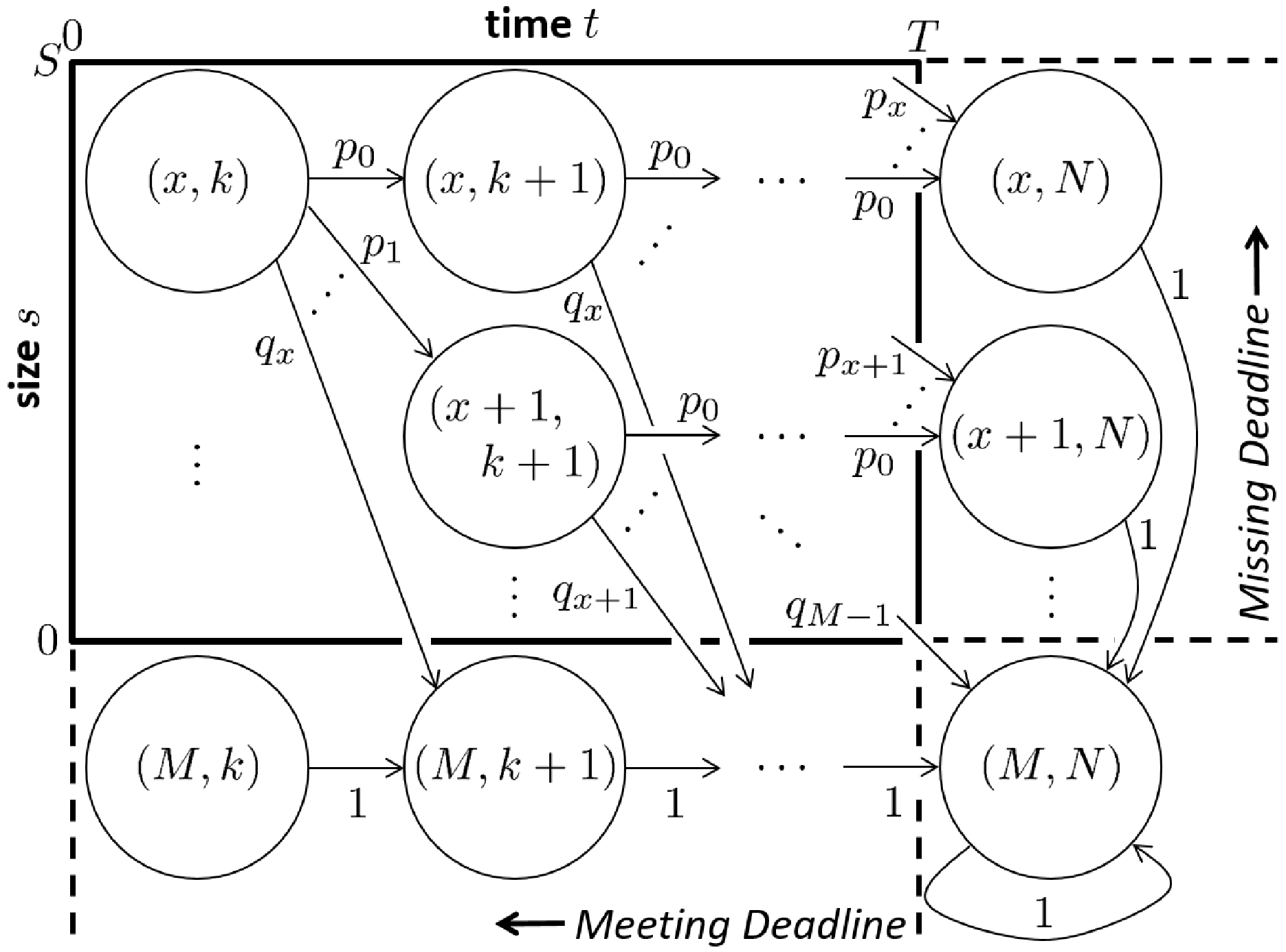}
\caption{\label{fig:TransitionProbs} The probable transitions in the stochastic
shortest path problem from the perspective of being in step $x$ at stage $k$ within the
size-time discretization in \FigRef{StateSpace}. Inside the size-time space, the (discrete) state
evolution is parameterized by progress probabilities $p_x$ and $q_x$, where influence
of control is captured by models to predict risks of missing the elastic deadline under
fixed-rate assumptions. Outside the size-time space, however, the
state evolves deterministically and without influence of control.}
\end{figure}

\subsection{From Risk Predictions to Transition Probabilites \label{ssec:TheProbs}}
The transition probabilities of a stochastic shortest path problem refer to the entries of
the size-$m$ collection of $n$-by-$n$ stochastic matrices as modeled in \equRef{SSPModel}.
Recall the discretization of \FigRef{StateSpace}, ordering
the $(M+1)$-by-$(N+1)$ grid of step-stage cells $(x,k)$ in correspondence with
 state index $i = 1, 2, \ldots, n$
via \equRef{StateSpace}. Also recall the correspondence between action index $a = 1, 2,
\ldots m$ and the series of mean elastic rates $R(a)$ ordered via \equRef{ActionSpace}.
\FigRef{TransitionProbs} illustrates the probable transitions from the perspective of
being in step $x < M$ at stage $k < N$, where those leaving states inside the
size-time space involve a length-$(M+1)$ probability vector
\begin{equation}
\label{equ:StepProbs}
\mat{p} =  \left[ \begin{array}{cccc} p_0 & p_1 & \cdots & p_M \end{array} \right]
\end{equation}
and the derived probability $q_x = \sum_{y = M-x}^M p_y$ for every $x$.
Before explaining all details, including how \equRef{StepProbs} is influenced by
admission control, we first discuss modeling the risk of missing the elastic
deadline under fixed-rate assumptions.

\subsubsection{Risk Predictions} Consider a duration of time (in seconds)
during which the mean elastic rate is held at $R$
(in megabits per second). Recall also that the discretization of the elastic
flow's continuous space involves size $\delta_S$ (in megabits per step) and
time $\delta_T$ (in seconds per stage). We assume that the elastic flow's
remaining size is non-increasing over time and that the number of steps
taken versus time is modeled by a given counting process \cite{Gal96:DisSP}
with mean rate $R/\delta_S$ (in steps per second), implying the probabilities
$\bar{p}_\ell$ for $\ell = 0, 1, 2, \ldots$ of progressing $\ell$ steps within
duration $\delta_T$ of any one stage.  For example, if the time between
successive steps is described by a sequence of independent and
identically-distributed exponential random variables with mean $\delta_S/R$
(in seconds per step), then the counting process is Poisson and thus
$$
\bar{p}_\ell =
\frac{e^{-\left(R\delta_T/\delta_S\right)}\left( R\delta_T/\delta_S \right)^\ell}{\ell!},
\quad  \ell = 0, 1, 2, \ldots .
$$
Also recognizing that starting a stage in step $x < M$ implies that all
additional steps $\ell\geq M-x$ result in completing the flow, the
probability vector of \equRef{StepProbs} is populated by
$$
p_y =
\left\{ \begin{array}{lcl}
\bar{p}_y & , & y = 0, 1, \ldots, M-1 \\[0.1in]
\sum_{\ell=M}^\infty \bar{p}_\ell & , & y = M
\end{array} \right.
$$
and, altogether,
the per-stage probabilities of progressing from step $x$ to step $x^\prime$ are
\begin{equation}
\label{equ:StepMatrix}
P_{x,x^\prime} = \left\{
\begin{array}{lcl}
0 & , & 0 \leq x^\prime < x \leq M \\
p_{x^\prime-x} & , & 0 \leq x \leq x^\prime < M \\
q_x & , & 0 \leq x < x^\prime = M \\
1 & , & x = x^\prime = M
\end{array}
\right.
\end{equation}
with $q_x = \sum_{y = M-x}^M p_y$.

Risk prediction under mean elastic rate $R$, accepting the choice of both the counting
process model and the size-time discretization, is now a matter of standard Markov chain
calculations \cite{BeT08:IntPr}. Observe that the per-stage
progress probabilities in \equRef{StepMatrix} express
an $(M+1)$-by-$(M+1)$ stochastic matrix upon indexing the row and column by current step
$x$ and subsequent step $x^\prime$, respectively:
$$
\mat{P} = \left[
\begin{array}{cccccc}
p_0    & p_1    & p_2    & \cdots & p_{M-1} & q_0 \\
0      & p_0    & p_1    & \cdots & p_{M-2} & q_1 \\
0      & 0      & p_0    & \cdots & p_{M-3} & q_2 \\
\vdots & \vdots & \ddots & \ddots & \vdots  & \vdots \\
0      & 0      & \cdots & 0      & p_0     & q_{M-1} \\
0      & 0      & \cdots & 0      & 0       & 1
\end{array}
\right] .
$$
Denote by $f_x(k)$ the probability of starting from step $x$ at stage $k$, where
$$
\mat{f}(k) = \left[ \begin{array}{cccc} f_0(k) & f_1(k) & \cdots & f_M(k) \end{array} \right]
$$
denotes the associated length-$(M+1)$ probability vector, conditioned
on starting from step $0$ at stage $0$ i.e., given
$$
\mat{f}(0) = \left[ \begin{array}{cccc} 1 & 0 & \cdots & 0 \end{array} \right] .
$$
It follows that the sequence of these stage-dependent probability vectors can be
calculated via matrix multiplication
$$
\mat{f}(k) = \mat{f}(k-1)\mat{P}, \quad k = 1, 2, \ldots, N.
$$
These vectors enable different types of prediction e.g.,
the mean and variance of progress (in steps) after $k$ stages is
$$
\theta(k) = \sum_{x=0}^M x f_x(k) \quad \mbox{and} \quad \sigma^2(k) = \sum_{x=0}^M
\left[x - \theta(k)\right]^2 f_x(k)
$$
respectively; the risk of \emph{not} reaching a specific step $x$ by stage $k$ is
probability $\sum_{y=0}^{x-1} f_y(k)$; the overall risk of missing the deadline
is probability $1 - f_M(N)$. Moreover, using stochastic matrix
$\mat{P}$ to calculate mean first-passage times or expected times to absorption can further
inform when a given milestone on remaining size will likely be reached. Finally,
using discretization parameters $\delta_S$ and $\delta_T$, progress measured in steps
versus stages (with its single standard deviation envelope) can approximate its
continuous counterpart measured in megabits versus seconds e.g., linearly
interpolate the sequence of size-time points $(s_k,\delta_Tk)$ with
$s_k = S - \delta_S\left[ \theta(k) \pm \sigma(k) \right]$.

\subsubsection{Transition Probabilities}
Consider first the states outside of the continuous size-time space:
cell $(M,N)$, or index $i = n$, is the designated terminal state, while all other cells
at step $M$ or stage $N$ correspond to meeting or missing, respectively, the deadline.
Altogether, it follows that each stochastic matrix $\mat{F}(a)$, regardless of action $a$,
assigns unit probability to
\begin{itemize}
\item the self-transition in cell $(M,N)$, which represents termination;
\item the transition from cell $(M,k)$ to cell $(M,k+1)$
for every stage $k < N$, which represents meeting the deadline with $N-k$ stages remaining; and
\item the transition from cell $(x,N)$ to cell $(M,N)$ for every step $x < M$, which represents
missing the deadline with $M-x$ steps remaining.
\end{itemize}
As a stochastic matrix, each such row with unit probability in one column must have zero
probability in all other columns.

It remains to consider the states inside of the continuous size-time space,
namely cells $(x,k)$ such that step $x < M$ and stage $k < N$.
\FigRef{TransitionProbs} assumes that stages evolve in succession
deterministically, no matter the applied action $a$, and
the upper-triangular matrix structure in \equRef{StepMatrix}
no matter the rate $R$, assumes it is impossible for step $x$ to decrease
from stage-to-stage. It follows that in each stochastic matrix $\mat{F}(a)$
only transitions from state $i \leftrightarrow (x,k)$ to state $j \leftrightarrow (x^\prime,k+1)$
such that step $x^\prime \geq x$ are probable, using matrix
$\mat{P}(a)$ under rate $R(a)$ to assign $F_{i,j}(a) = P_{x,x^\prime}(a)$.

It is worth noting ways in which our probabilistic model admits generalization.
Observe that no part of the scheme by which the progress probabilities
in $\mat{P}(a)$ map into the transition probabilities in $\mat{F}(a)$ require each
row of $\mat{P}(a)$ to stem from a counting process common to all steps.
Indeed, matrix $\mat{P}(a)$ could result from an underlying
Markov process with step-dependent transition rates, say if the notion of step is
discretizing observed quantities other than just the elastic flow's remaining size.
Also observe that no part of the scheme depends on progress probabilities being the
same for all stages, say if an action's elastic rate is expected to vary with stage
because link bandwidth or the action's inelastic load is known to predictably vary
with time.

\subsection{From Reward Functions to Transitions Costs \label{ssec:TheCosts}}
The cost matrices as modeled in \equRef{SSPModel}
exhibit the same organization as the stochastic matrices
just discussed. That is, the $n$ states again correspond to the
discretization of \FigRef{StateSpace} via \equRef{StateSpace} and the $m$ actions
again correspond to the mean elastic rates via \equRef{ActionSpace}. The setup
of \FigRef{MainSetup} suggests a tension between maintaining negligible risk
of missing the elastic deadline yet sustaining the most profitable inelastic traffic.
This tension will be quantified by transition cost matrices of the form
\begin{equation}
\label{equ:DualObjective}
\mat{G}(a) = \mat{G}^E(a) + \lambda^I \mat{G}^I(a)
\end{equation}
where $\mat{G}^E$ and $\mat{G}^I$ reflect, respectively, the per-stage
elastic and inelastic rewards (in utils), while $\lambda^I \geq 0$ parameterizes
the optimized balance between the competing objectives.

\subsubsection{Reward Functions}
Consider first the elastic reward function $V^E(t)$ versus elapsed time $t$, which we
assume takes arbitrary positive values until deadline $T$ but must be zero beyond $T$.
It represents the one-time reward (in utils) incurred upon completion of the elastic flow. The
omission of remaining size $s$ reflects the assumption that no reward is incurred prior to
completion, even if that size has been considerably diminished by the deadline.
That we assume no reward for late completion views $T$ as a hard deadline, adopting a
``better-never-than-late'' posture in the elastic objective. However, permitting arbitrary values
until time $T$ can capture soft deadlines (via an abrupt decrease in
reward) or other forms of preferred completion times. An elementary choice is
$V^E(t) = 1$ over $[0,T]$, which in expectation renders the probability of
meeting the deadline.

The inelastic reward function $V^I(a)$ depends only on the applied action $a$, with non-negative
values representing the reward rate (in utils per second) upon admission of the associated subset
of inelastic flows. Recalling from \equRef{ActionSpace} that actions are ordered by non-decreasing
inelastic load, reward rates will be similarly ordered for worthwhile admissions:
\begin{equation}
\label{equ:InelasticRewards}
0 = V^I(1) < V^I(2) \leq \cdots \leq V^I(m).
\end{equation}
Lastly, it will be our convention to calibrate all inelastic reward rates with respect
to the elastic reward function, meaning to proportionally scale $V^I$ such that
$$
V^I(m) = \frac{V^E(T)}{T},
$$
which equates the total reward accrued from the length-$T$ admission of
the most-profitable inelastic subset with the
one-time reward of completing the elastic flow at exactly the deadline.
Thereafter, the balance between elastic and inelastic objectives remains tunable by choice
of parameter $\lambda^I$ in \equRef{DualObjective}.

\subsubsection{Transition Costs}
Assigning costs, or negative reward, to zero probability transitions is
superfluous, so only the probable transitions represented in \FigRef{TransitionProbs}
need consideration. In the elastic cost matrices $\mat{G}^E(a)$, the only non-zero
entries under every action $a$ are the negated one-time rewards for transitions
that represent completing the elastic flow within deadline. These transitions are from
state $i \leftrightarrow (x,k)$ to state $j \leftrightarrow (M,k+1)$
such that step $x < M$ and stage $k < N$, assigning cost $G^E_{i,j}(a) = -V^E(t)$ with
time $t = (k+1)\delta_T$ corresponding to the end of stage $k$. In the inelastic cost
matrices $\mat{G}^I(a)$, the only zero entries under every action $a$ are for transitions
from state $i \leftrightarrow (x,N)$ to state $j = n$, representing termination. All other
probable transitions are assigned the cost $G^I_{i,j}(a) = -V^I(a)\delta_T$.

\subsection{On Policy Optimization, Evaluation and Implementation \label{ssec:TheAlgs}}
The theory of discrete-time dynamic programming for $n$-state, $m$-action stochastic
shortest path problems is rich, offering firm technical conditions that guarantee
(i)~some choice of stationary policies, or functions of the form
${\mu:\{ 1,2,\ldots,n \} \rightarrow \{ 1,2,\ldots,m \}}$,
minimizes the expected sum-total cost $J_i$ (in negated utils) assuming initial
state $i$ and (ii)~convergence of
the standard algorithms (e.g., value iteration, policy iteration)
by which to evaluate the length-$n$ vector $\mat{J}(\mu)$ for any given policy as well as
obtain an optimal policy $\mu^* =\arg\min_\mu \mat{J}(\mu)$. For our formulation,
in particular, these guarantees hold by virtue of the deterministic transitions
that represent monotonically increasing stages until termination.

The key output of these dynamic programming algorithms is a $n$-by-$m$
matrix $\mat{Q}(\mu)$ whose row-$i$, column-$a$ entry is the expected
cost-to-go starting from state $i$,
applying action $a$ and employing the policy $\mu$ therafter.
In policy optimization, the resulting matrix $\mat{Q}$ implies the function
$\mu^*$ by satisfying for every $i$
$$
\mu^*(i) = \arg \min_{a \in \{ 1,2,\ldots,m \}} Q_{i,a}.
$$
Policy evaluation reduces simply to $J_i = Q_{i,\mu(i)}$. The separation of
single-stage costs into elastic and inelastic components in
\equRef{DualObjective} is preserved in the expected sum-total cost i.e.,
for any policy $\mu$
$$
\mat{J}(\mu) = \mat{J}^E(\mu) + \lambda^I \mat{J}^I(\mu).
$$
However, the output of policy optimization against a given model
$\mat{F}$ and $\mat{G}$ will not explicitly separate $\mat{J}(\mu^*)$ into
the elastic and inelastic components. Determining this separation requires
at least one additional call to the algorithms  e.g., evaluate the
optimal policy against model $\mat{F}$ and $\mat{G}^I$ to obtain $\mat{J}^I$,
then obtaining $\mat{J}^E = \mat{J} - \lambda^I\mat{J}^I$.

Policy implementation aligns the manner in which the discretized state process
unfolds in stages to the periodic opportunity for admission control
based upon the elastic flow's status $z_t$ presented in \FigRef{MainSetup}. Specifically, upon
entering stage $k$ at time $t = k\delta_T$ and identifying step $x$ from the
interval $\left( S-(x+1)\delta_S,S-x\delta_S \right]$ that contains the remaining size
$s$, the corresponding state index $i \leftrightarrow (x,k)$  feeds the policy
by which control $u_t = \mu(i)$ is invoked. The associated mean elastic rate $R(u_t)$
is held until time $t = (k+1)\delta_T$ at which point the revised size $s^\prime$
identifies the subsequent step $x^\prime$. The transition from cell $(x,k)$ to
cell $(x^\prime,k+1) \leftrightarrow j $, realizing the single-stage
reward $-G_{i,j}(u_t)$, simultaneously ends the current control period
and begins the next.

It is worth noting that all
risk predictions discussed in \sSecRef{TheProbs} on a per-action
basis can be made for the controlled process.
This requires first extracting the length-$N$ sequence of $(M+1)$-by-$(M+1)$
stochastic matrices that collectively represent the (albeit now stage-dependent)
progress probabilities under given policy $\mu$.
More specifically, for each stage  $k = 0, 1, \ldots, N-1$, define the
(upper-triangular) matrix $\mat{P}^\mu(k)$ by assigning to each row $x$
and column $x^\prime$ the transition probability
$$
P^\mu_{x,x^\prime}(k) = F_{i,j}(\mu(i)),
$$
where $i \leftrightarrow (x,k)$ and $j \leftrightarrow (x^\prime,k+1)$.
Policy-dependent risk predictions analogous to those discussed in
\sSecRef{TheProbs} are now possible e.g.,
calculating the stage-dependent probability vectors via
matrix multiplication
$$
\mat{f}(k) = \mat{f}(k-1)\mat{P}^\mu(k-1), \quad k = 1, 2, \ldots, N.
$$

Finally, note also that an alternative representation of stationary policy $\mu$
over the set of step-stage pairs $(x,k)$, by virtue
of the deterministic evolution over successive stages, is as a sequence of
stage-dependent policies $\left( \mu_0, \mu_1, \ldots, \mu_{N-1}\right)$,
each component a function ${\mu_k:\{ 0,1,\ldots,M \} \rightarrow \{1,2,\ldots,m\}}$
over only steps $x$. We have approached policy optimization under an
infinite-horizon criterion because, for problems with a finite control space,
the associated algorithms are generally more computationally
efficient than dynamic programming methods to solve the finite-horizon counterpart.
However, harnessing these infinite-horizon efficiencies
in our setup comes with the discretized approximation of system dynamics as well as the expense of
including stage $k$ in the state space. Indeed, posing and analyzing alternative finite-horizon formulations
is among the items for future research suggested in \SecRef{Conclu}.

\section{Examples and Experiments \label{sec:NomAnal}}
This section employs the formulation of \SecRef{ProbForm} to analyze a specific
scenario within the setup of \FigRef{MainSetup}. The primary goal is to develop
insights into the performance gains achievable by our admission control policies in
comparison to uncontrolled (i.e., fixed admission) policies. While a sensible
tradeoff between reducing elastic risk and sustaining inelastic traffic is
demonstrated in all cases, the curves achieved via our controlled policies are
consistently superior. A secondary goal is to illustrate the interplay of key
structural parameters (e.g., discretization of the size-time space, admissible
subsets of inelastic flows) in forming a tractable stochastic shortest path problem.
Our experiments will also visualize the risk predictions that lead to
stochastic matrices $\mat{F}$ as well as examine the impact of
alternative reward functions that lead to cost matrices $\mat{G}$.

\subsection{Scenario Specifics and Baseline Results \label{ssec:BaselineResults}}
Consider a 200~Mbps link dedicated to 51 different flows, the primary elastic flow
being the transfer of a 30~GB file to be completed within 30 minutes in contention with
fifty equally-profitable inelastic flows: twenty-five for VoIP traffic with
per-flow load of 0.1~Mbps and twenty-five for video traffic with per-flow
load of 3~Mbps. We thus have parameters ${B = 200}$,
$S = 240,000$ and $T = 1800$ in the setup of \FigRef{MainSetup} with $m\geq 2$
candidate actions associated to elastic rates $200 = R(1) > R(m) \geq 122.5$, but
the exact control space remains as yet unspecified because the collection of all
subsets of inelastic flows is clearly too large. We also take the Poisson counting
process to model the risk predictions, as discussed in \sSecRef{TheProbs}, as well
as the elastic reward function to indicate meeting the deadline, as discussed
in \sSecRef{TheCosts}.

\begin{figure}[t]
\centering
\vspace{0.1in}
\includegraphics[width=0.475\columnwidth]{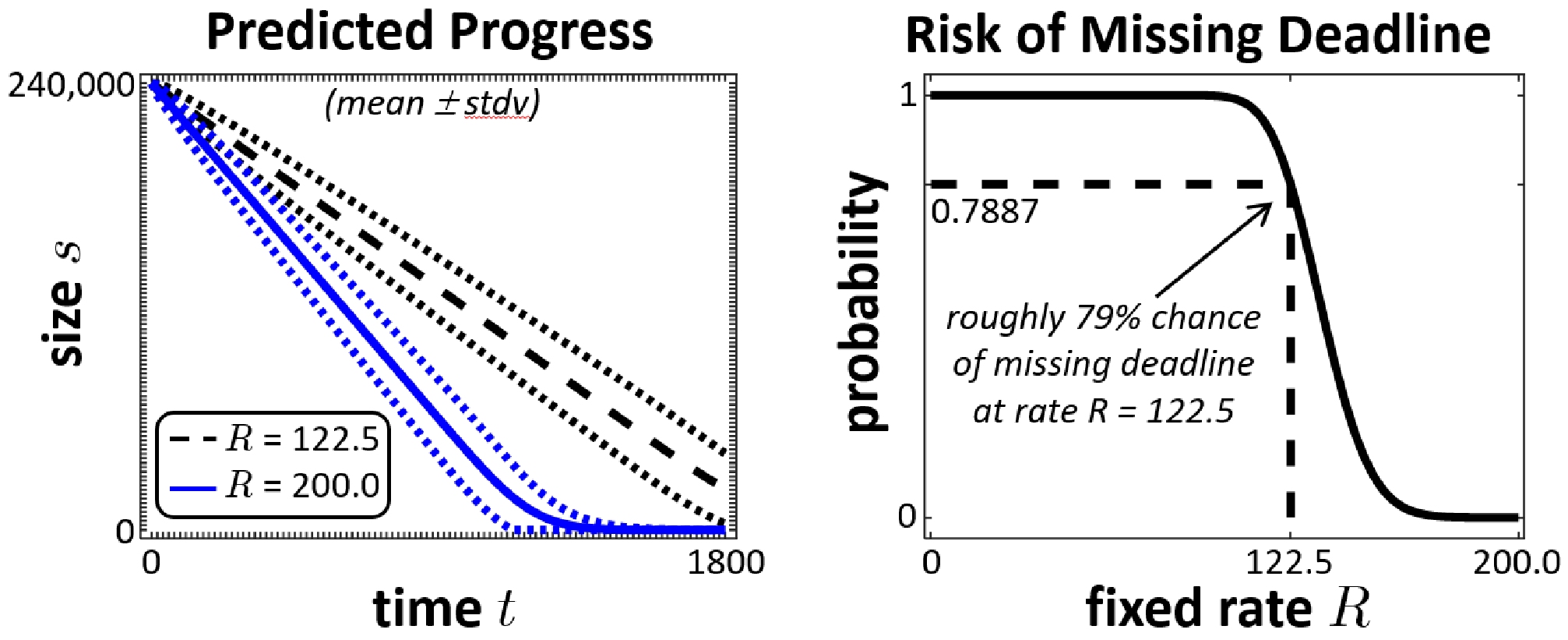} \\
{\footnotesize (a)~Elastic Traffic Characterization and the State Space ($M = N = 100$)} \\[0.2in]
\includegraphics[width=0.475\columnwidth]{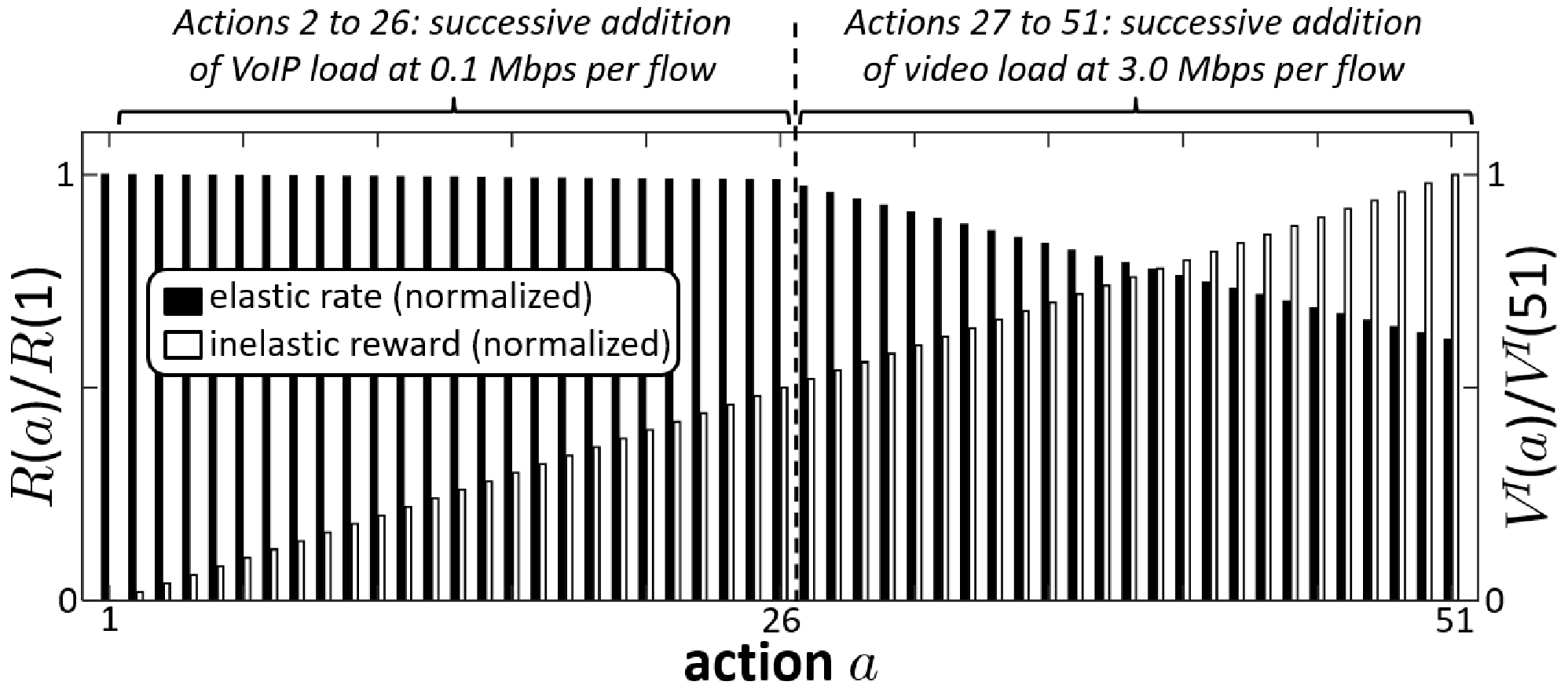} \\
{\footnotesize (b)~Inelastic Traffic Characterization and the Control Space ($m = 51$)}
\caption{\label{fig:BaselineScenarioSetup} From scenario parameters to choice of model structure,
risk predictions and reward functions given a 200~Mbps link ($B = 200$) dedicated to 51 different flows,
the primary elastic flow being the transfer of a 30~GB ($S = 240,000$) file to be completed within 30
minutes ($T = 1800$) in contention with fifty equally-profitable inelastic flows: twenty-five for VoIP
traffic with per-flow load of 0.1~Mbps and twenty-five for video traffic with per-flow
load of 3~Mbps. (a)~The size-time state space is derived from attributes of the elastic flow and
discretized into a 101-by-101 grid of step-stage cells, taking a Poisson counting process to model
the risk predictions as described in \sSecRef{TheProbs}. (b)~The size-51 control space employs a
scheme that successively adds inelastic flows as action index $a$ increases, taking the elastic reward function
to indicate meeting the deadline and calibrating the inelastic reward rates as described in \sSecRef{TheCosts}.}
\end{figure}

\FigRef{BaselineScenarioSetup}(a) characterizes the elastic traffic for
this scenario upon choosing discretization parameters $M = N = 100$, yielding
$\delta_S = 2400$ and $\delta_T = 18$. The left plot predicts, for both the minimum
and maximum elastic rates of 122.5 and 200.0, respectively, the elastic flow's progress
(in megabits versus seconds, along with its standard-deviation envelope).
The right plot predicts the overall risk, or probability of missing the
elastic deadline, versus the (fixed) rate $R$ in $[0,B]$. \FigRef{BaselineScenarioSetup}(b)
characterizes the inelastic traffic upon adopting the following subset construction scheme,
yielding a control space that scales linearly with the number of inelastic flows:
given $m-1$ distinct inelastic flows, each $\ell$th flow associated to offered load
$\bar{L}_\ell$ (in megabits per second) and reward rate $\bar{V}_\ell$ (in utils per second),
order the indexing such that
$$
\frac{\bar{V}_1}{\bar{L}_1} \geq \frac{\bar{V}_2}{\bar{L}_2} \geq \cdots \geq
\frac{\bar{V}_{m-1}}{\bar{L}_{m-1}}
$$
and associate each action $a$ to the admission of flows ${\ell < a}$, aggregating
the offered load and reward rate additively i.e., for $a = 1,2, \ldots, m$, let
\begin{equation}
\label{equ:aggregation}
R(a) = B - \sum_{\ell=1}^{a-1} \bar{L}_\ell \quad \mbox{and} \quad V^I(a) \propto \sum_{\ell=1}^{a-1} \bar{V}_\ell.
\end{equation}
In our specific scenario, with all fifty inelastic flows being equally profitable,
this results in $m = 51$ actions, successively adding the VoIP flows
in actions 2 to 26 and the video flows in actions 27 to 51. The inelastic reward
rates are thus proportional to the fraction of admitted flows, which %because $V^E(T) = 1$
renders ${V^I(a) = \frac{1}{T}(a-1)/(m-1)}$ upon calibration.

\begin{figure}[t!]
\centering
\vspace{0.1in}
\includegraphics[width=0.45\columnwidth]{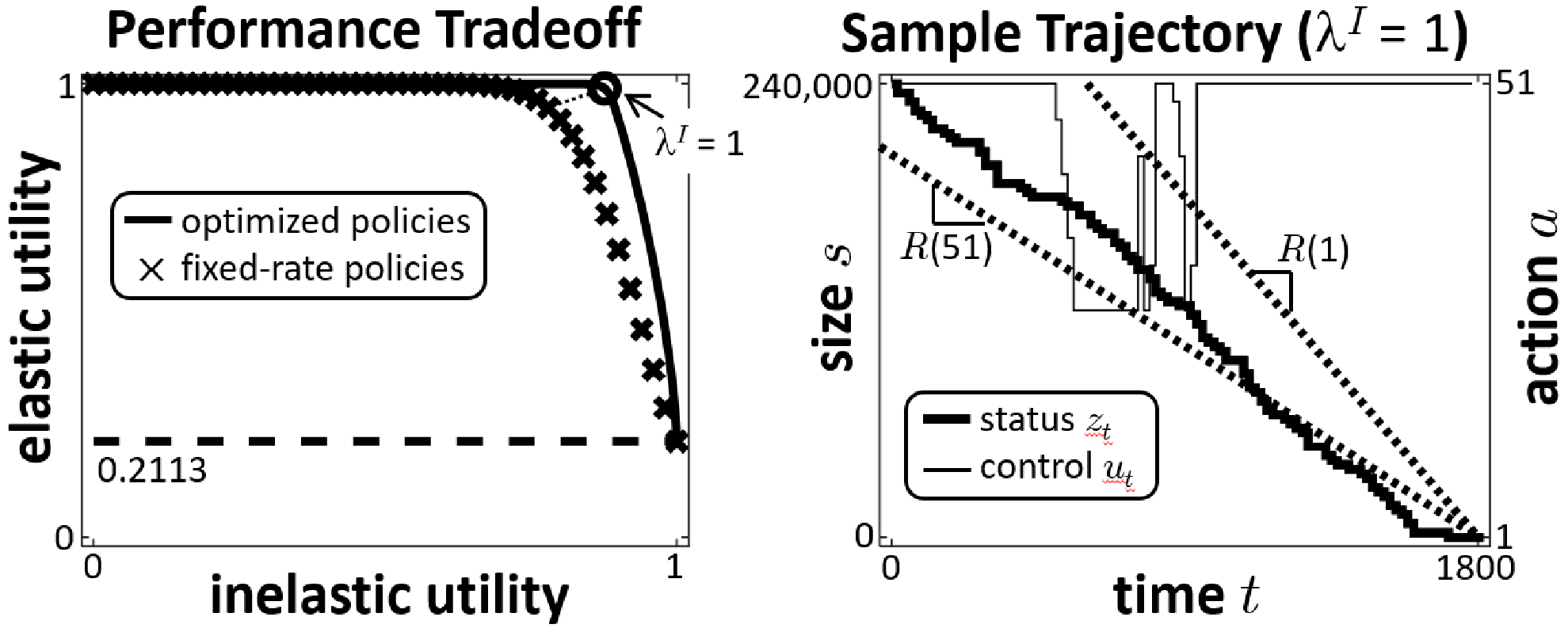} \\
{\footnotesize (a)~Performance and Sample Trajectory of the Controlled Process} \\[0.2in]
\includegraphics[width=0.475\columnwidth]{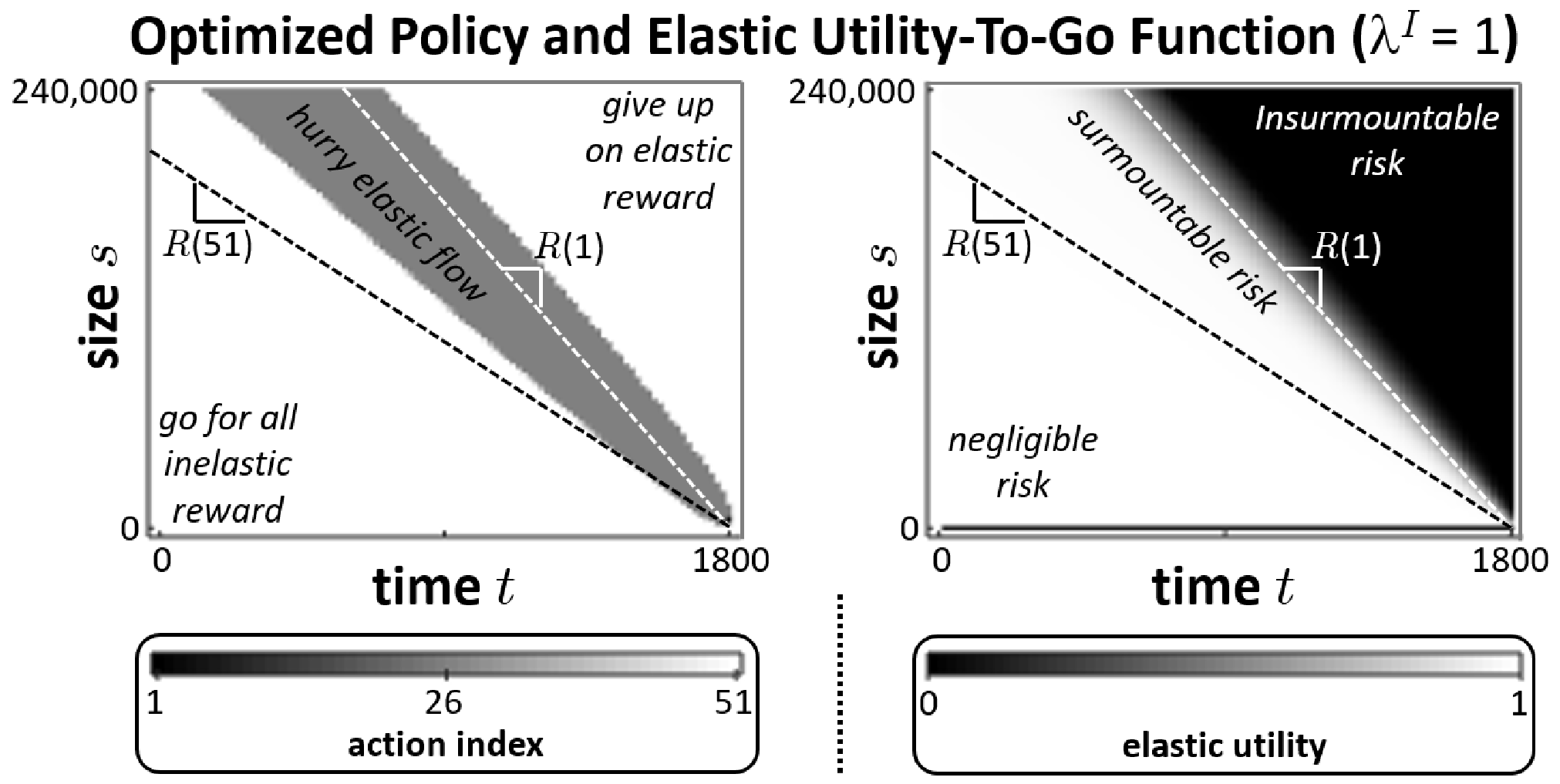} \\[0.1in]
\includegraphics[width=0.475\columnwidth]{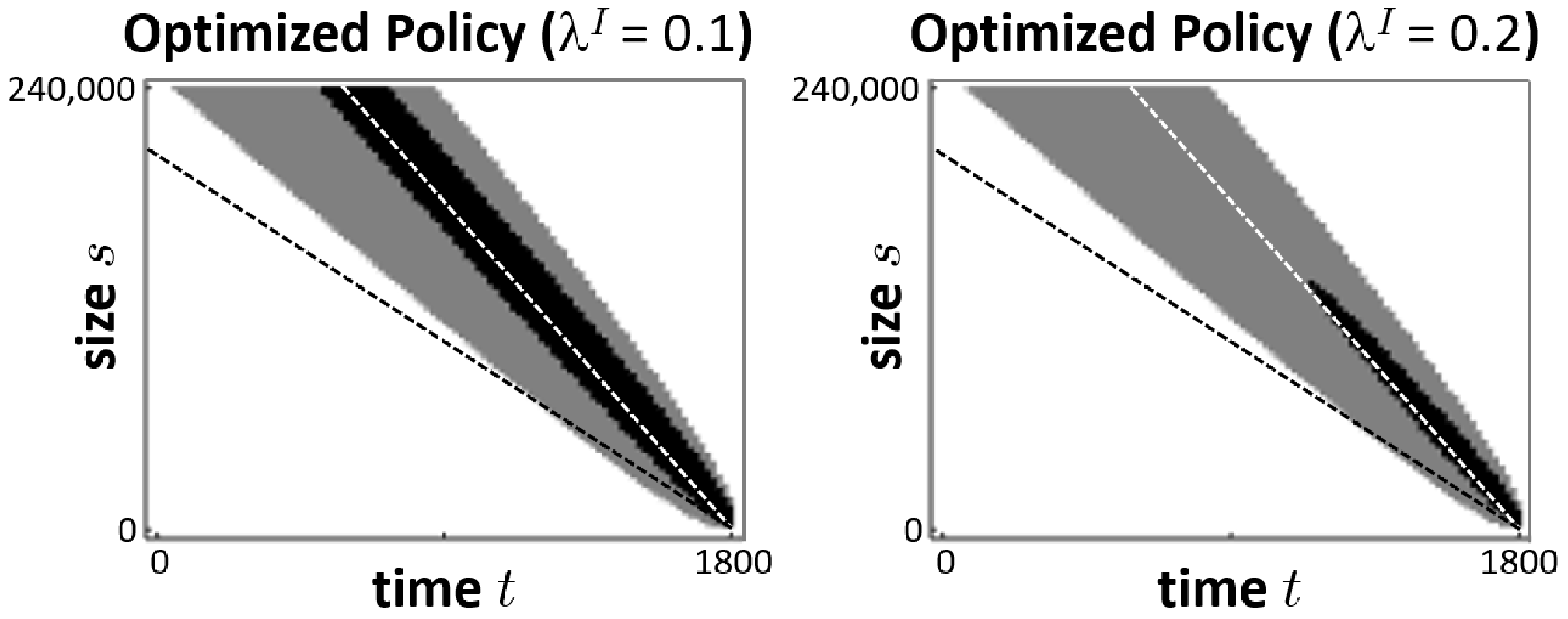} \\[0.1in]
\includegraphics[width=0.475\columnwidth]{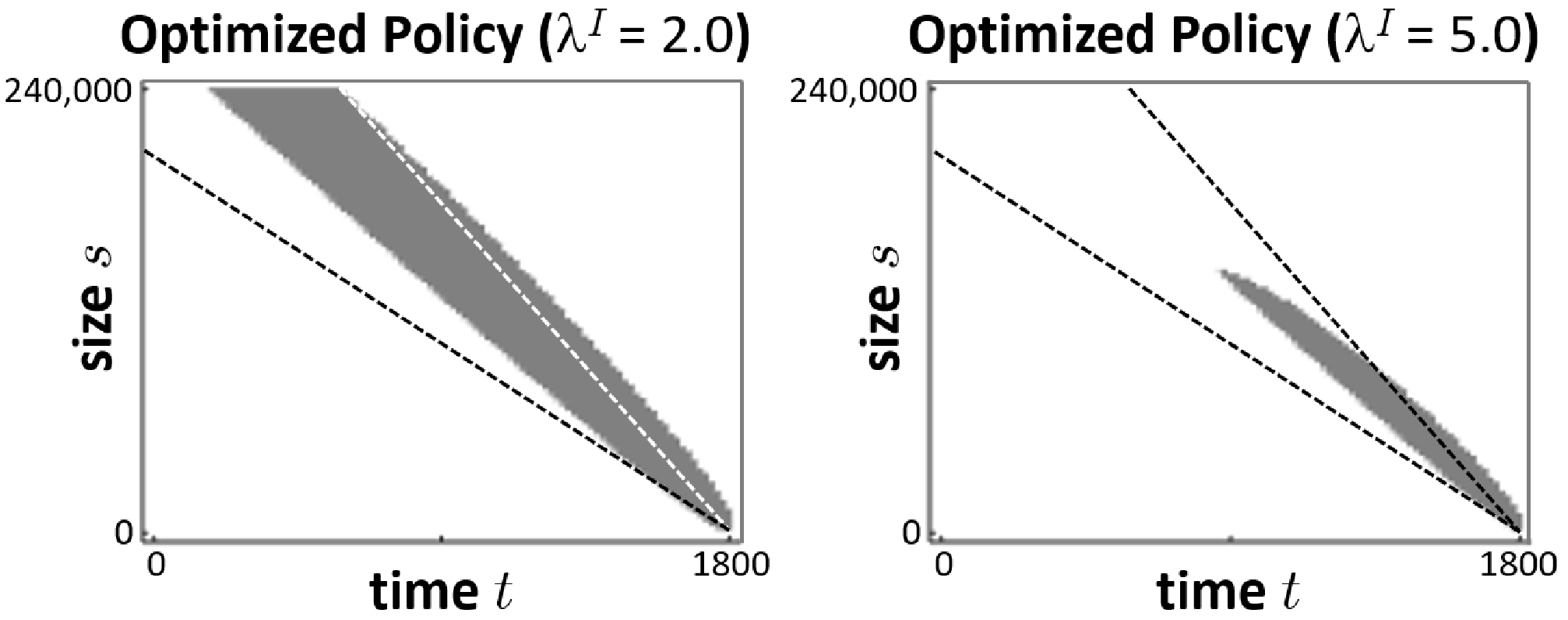} \\
{\footnotesize (b)~Behavioral Details of the Optimized Admission Controller}
\caption{\label{fig:BaselineScenarioResults} Baseline results of the stochastic shortest path analysis for the
scenario summarized in \FigRef{BaselineScenarioSetup}. (a)~The optimized
performance tradeoff corresponds to sweeping parameter $\lambda^I$ from small to large, in each case calling upon
policy optimization with elastic and inelastic utility given, respectively,
by $-J^E_1(\mu^*)$ and $-J^I_1(\mu^*)$. The fixed-rate tradeoff calls only upon policy evaluation, twice per
action $a$ to obtain both the elastic and inelastic utilities with $\mu(i) = a$ for all $i$. Also shown is
a typical realization of the controlled process, under the policy optimized for $\lambda^I = 1$. The elastic
flow is hurried only during the second quarter of the time horizon and otherwise admits all inelastic flows,
still meeting the deadline with minutes to spare. (b)~The policy and its elastic utility-to-go function for the case
with $\lambda^I = 1$, which denies inelastic flows to hurry the elastic flow only in states where risk is deemed
surmountable and otherwise admits all inelastic flows because risk is deemed either negligible or
insurmountable.
Also shown are policies for select values of $\lambda^I$, demonstrating its influence
on both the exact partitioning of the state space as well as what actions constitute hurrying the elastic flow.
Small $\lambda^I$ makes even the little additional load of the VoIP
flows potentially worth shedding, while large $\lambda^I$ will shed only the video flows and only when sufficiently close
to completion as the deadline nears.}
\end{figure}

\FigRef{BaselineScenarioResults} summarizes our analysis of the stochastic shortest path problem
stemming from the setup in \FigRef{BaselineScenarioSetup}.
All other things equal, we expect performance of the optimized policy, with the freedom to
dynamically regulate the inelastic load, to never do worse than any one of the $m$ uncontrolled
(i.e., fixed-load) policies. \FigRef{BaselineScenarioResults}(a)
illustrates this explicitly, showing the optimized tradeoff between meeting elastic and inelastic
objectives is indeed superior to the tradeoff exhibited by the set of fixed-rate policies.
\FigRef{BaselineScenarioResults}(b) provides more insight into the nature of these optimized
policies, detailing both the policy and the elastic utility-to-go for
different values of $\lambda^I$. In all cases the size-time space is essentially
partitioned into two behaviors but the specifics of that partitioning do vary with $\lambda^I$.
One behavior admits no video flows in the interest of hurrying the elastic flow, typically still
admitting the VoIP flows as they capture half of the inelastic reward
with relatively little impact to the elastic rate; this partition of the size-time space concentrates along the
mean progress line that intersects the point $(0,T)$ under maximum elastic rate $R(1) = B$. The other
behavior admits all flows but for reasons that depend upon the elastic
flow's status: the lower-left portion of the size-time space deems the status as sufficiently ahead-of-schedule
to safely go for all inelastic reward, whereas the upper-right portion deems the status as sufficiently behind
schedule to proactively give up on the elastic reward.

\subsection{On the Choice of Model Structure}
A key choice in our stochastic shortest path formulation is model structure,
collectively the number of states ${n = (M+1)(N+1)}$ in the discretization of the elastic
flow's size-time space as well as the number of actions $m$ associated to the admissible
subsets of inelastic flows. This subsection revisits the scenario of \sSecRef{BaselineResults}
and reports on select experiments with different choices of model structure. The primary
consideration is complexity in representation and computation, as the model involves
two size-$m$ collections of $n$-by-$n$ matrices
and the iterative algorithms are (roughly) order $mn^2$. The fidelity of the resulting
admission controller is also at stake, however.

\begin{figure}[t]
\centering
\vspace{0.1in}
\includegraphics[width=0.475\columnwidth]{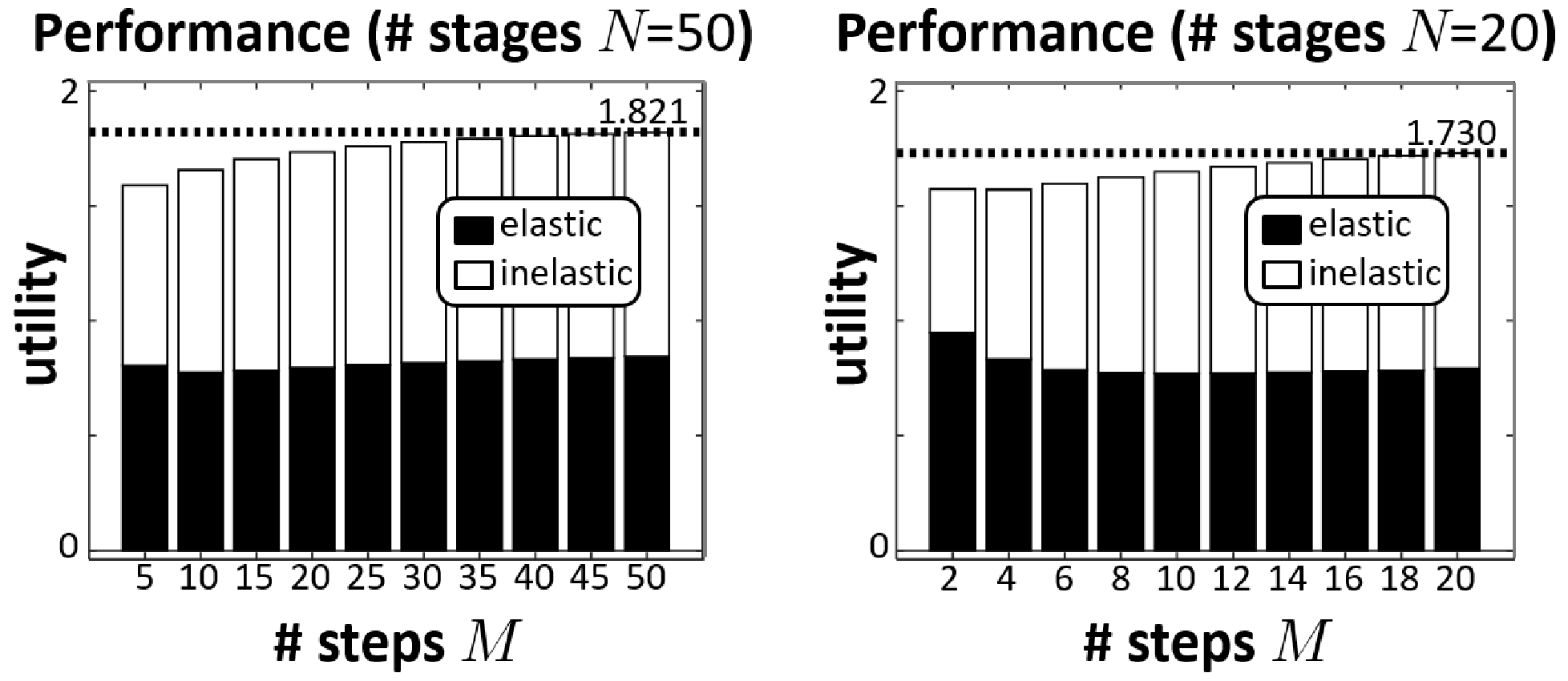} \\
{\footnotesize (a)~Performance for Different Choices of State Space (with $\lambda^I = 1$)} \\[0.2in]
\includegraphics[width=0.475\columnwidth]{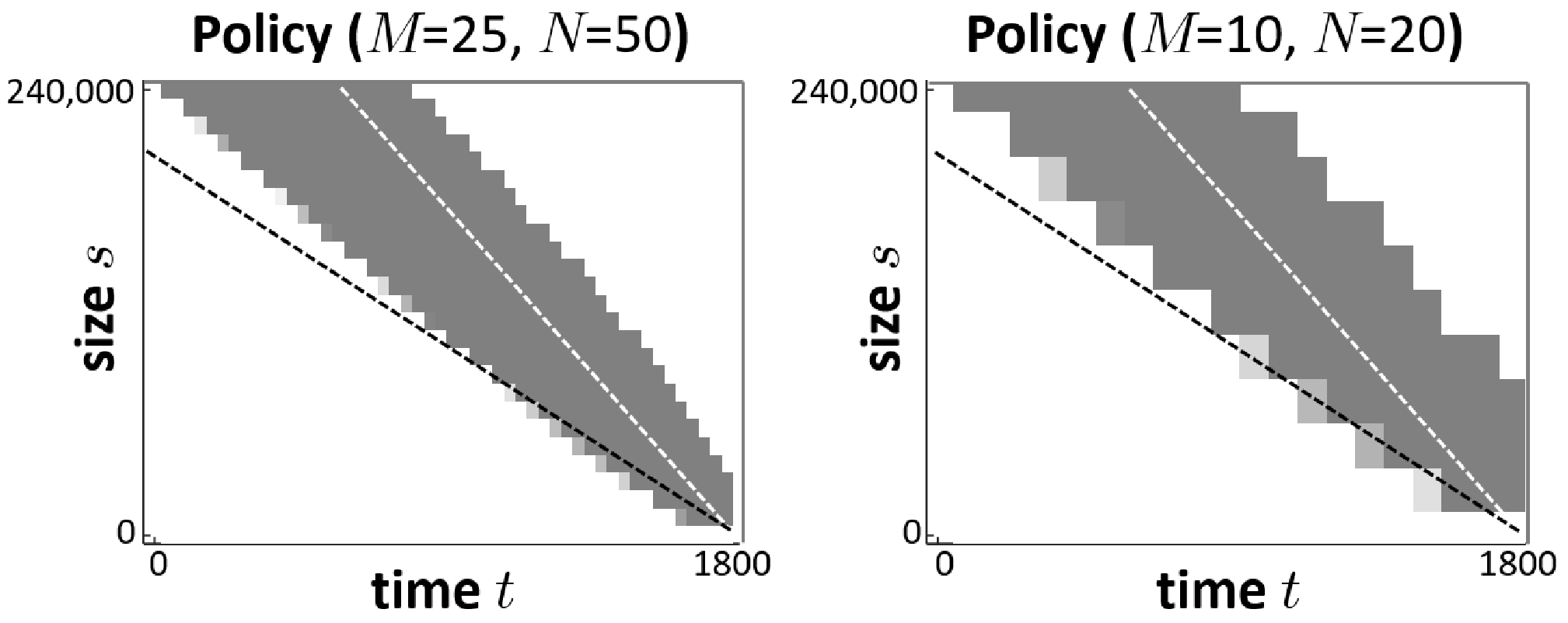} \\
{\footnotesize (b)~Policies for Different Choices of State Space (with $\lambda^I = 1$)}
\caption{\label{fig:ModelStructureState} Results of the stochastic shortest path analysis for the
scenario summarized in \FigRef{BaselineScenarioSetup} but with smaller numbers of states than the
choice of $M = N = 100$ in \FigRef{BaselineScenarioResults}, demonstrating
(a)~diminishing utility by virtue of the controller's reduced agility as (b)~the state space
is coarsened.}
\end{figure}

Recall that the number of stages $N$
is inversely proportional to the duration of time $\delta_T$ between successive control
decisions, while the number of steps $M$ is inversely proportional to the discernible
change in size $\delta_S$ within the feedback signal. Coarsening either, all other
things equal, restricts the agility of the associated admission controller.
\FigRef{ModelStructureState}(a) compares the utility of the optimized policy
(holding $\lambda^I = 1$) for smaller choices of state space. There are two plots,
holding the number of stages fixed at $N = 50$ and $N = 20$, respectively, while
letting the number of steps $M$ increase from $N/10$ up to $N$ in increments
of $N /10$. Each plot shows utility decreasing as $M$ decreases, and comparing between
plots (also noting that utility was 1.866 with $N = M = 100$ in \FigRef{BaselineScenarioResults}) similarly
shows diminished utility as $N$ decreases. \FigRef{ModelStructureState}(b) shows for each choice
of $N$ the policy associated with $M = N/2$, which upon comparison with the policies in
\FigRef{BaselineScenarioResults}(b) conveys the coarsened states and, in turn, the
less agile admission control underlying the diminished utility.

\begin{figure}[t]
\centering
\vspace{0.1in}
\includegraphics[width=0.475\columnwidth]{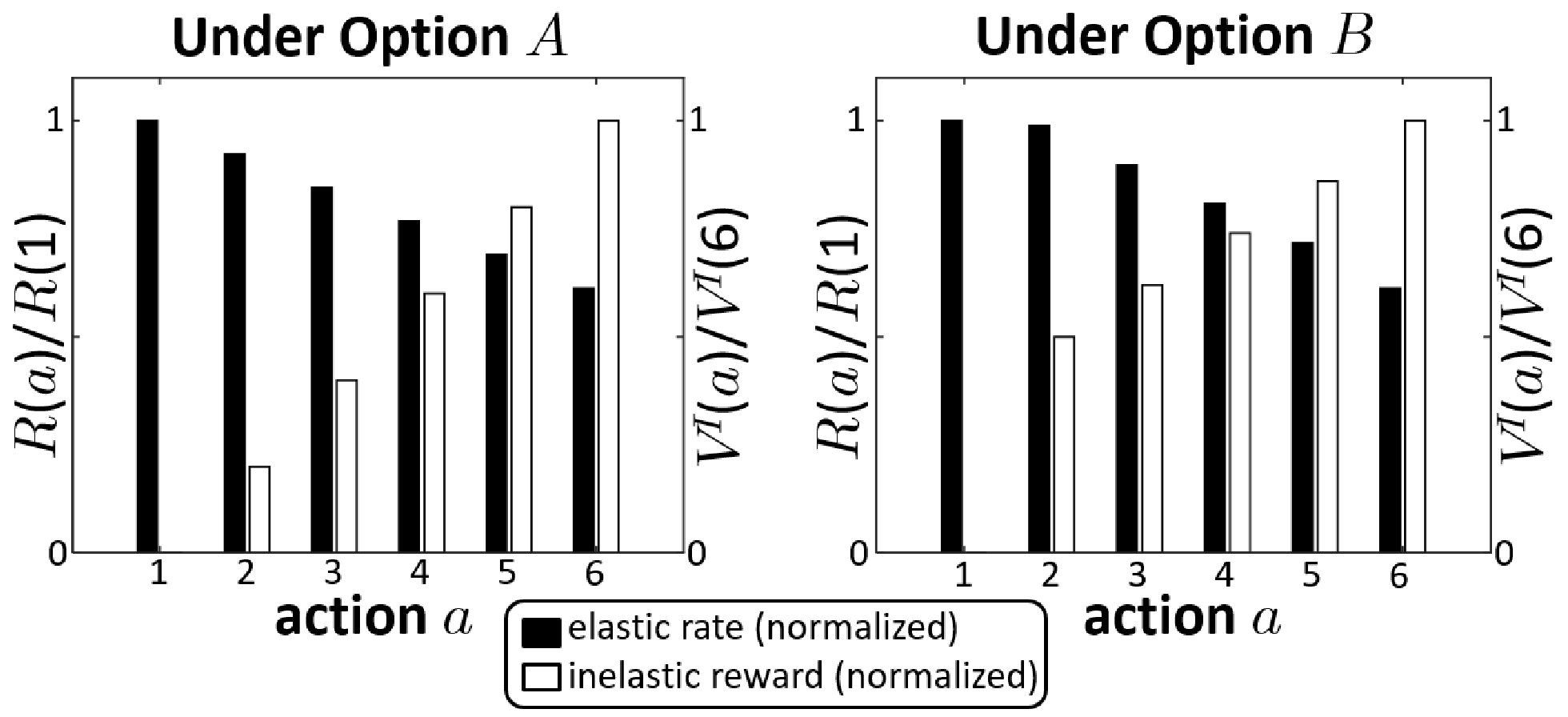} \\
{\footnotesize (a)~Two Options for the Reduced Control Space (with $m=6$)} \\[0.2in]
\hspace*{-0.1in}\includegraphics[width=0.45\columnwidth]{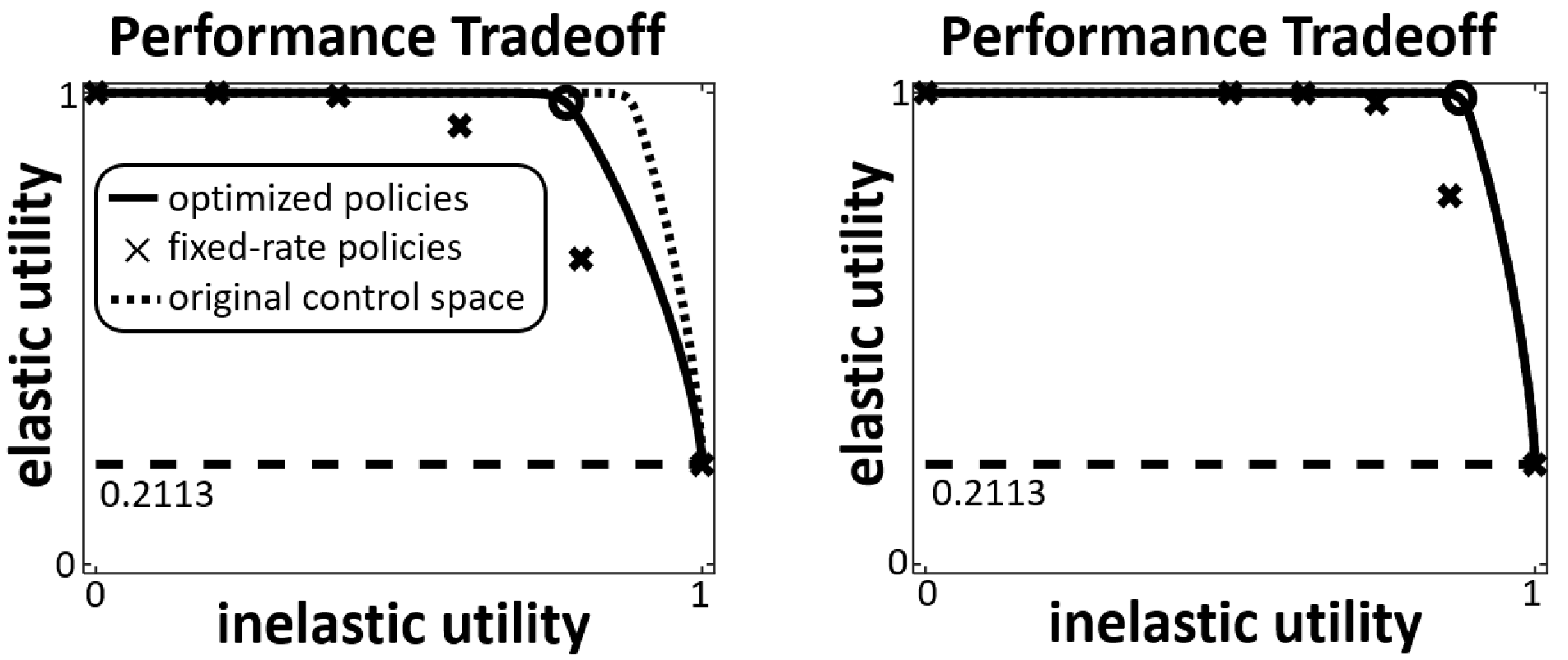} \\[0.1in]
\hspace*{-0.2in}\includegraphics[width=0.45\columnwidth]{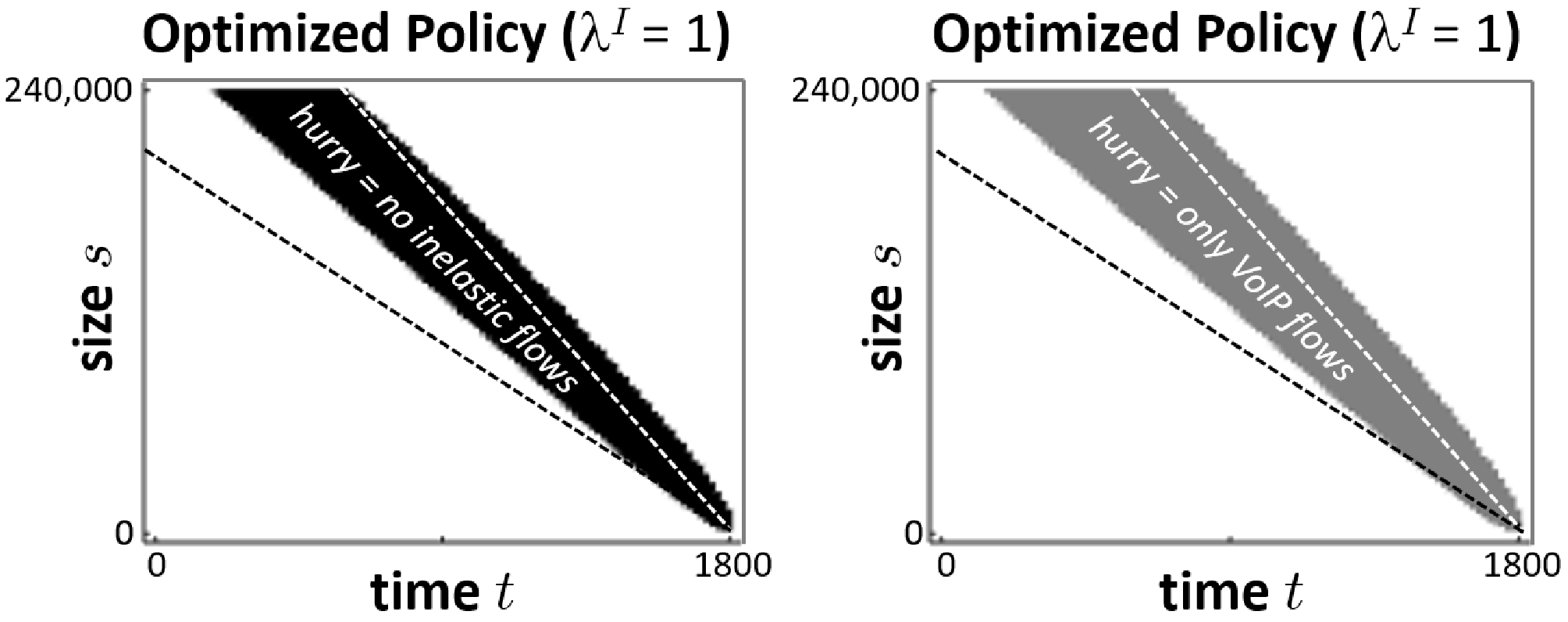} \\
{\footnotesize (b)~Behavioral Details of the Admission Controller Under Each Option}
\caption{\label{fig:ModelStructureControl} Results for the
scenario summarized in \FigRef{BaselineScenarioSetup} but with only $m = 6$ actions comprising the
control space (as opposed to the choice of $m = 51$  in \FigRef{BaselineScenarioResults}). (a)~The
reduced control space is constructed under two different options for dividing the fifty
inelastic flows into five groups: option $A$ (left column) mixes the VoIP and video flows to render groups of
equal load/reward, while option $B$ (right column) assigns all VoIP flows to one group giving it low load and
high reward relative to the four groups of only video flows. (b)~Only option $A$ exhibits noticeable
degradation from the performance in \FigRef{BaselineScenarioResults}, demonstrating that the
impact of a reduced control space can depend heavily on scenario specifics.}
\end{figure}

Reducing the number of actions $m$ also limits control performance but somewhat differently than
coarsened states. Smaller $m$ generally restricts the controller's flexibility,
within a given agility resulting in a smaller number of options to balance between the elastic
and inelastic objectives. However, the extent to which optimized performance is
impacted by this diminished flexibility is more dependant on scenario specifics than is the case with
reduced agility. To illustrate the point, impose the requirement that the scenario of
\sSecRef{BaselineResults} be controlled using only $m = 6$ actions. Consider two different options
for meeting this requirement: option $A$ divides the fifty inelastic flows into five
equally-profitable groups, each containing five VoIP flows and five video flows; option $B$
divides the inelastic flows non-uniformly, where one group contains all twenty-five VoIP flows,
another group contains seven video flows and the other three groups each contain six video flows.
Each option initializes the subset construction scheme described in \equRef{aggregation} differently:
$$
\begin{array}{ll}
\mbox{Option $A$:} & (\bar{L}_\ell,\bar{V}_\ell) = (15.5,10) \mbox{ for } \ell = 1,2, \ldots, 5 \\[0.1in]
\mbox{Option $B$:} & (\bar{L}_1,\bar{V}_1) = (2.5,25), (\bar{L}_2,\bar{V}_2) = (21,7) \mbox{ and } \\
& (\bar{L}_\ell,\bar{V}_\ell) = (18,6) \mbox{ for } \ell = 3, 4, 5
\end{array}
$$
\FigRef{ModelStructureControl} compares the two options for reducing the control space to
just $m = 6$ actions, both in terms of performance (via sweeping parameter $\lambda^I$)
as well as the policy (with $\lambda^I = 1$). Only option $A$ exhibits noticeable degradation
from the performance in \FigRef{BaselineScenarioResults}. This is because the smaller control
space constructed under option $B$ includes the inelastic subsets that are typically admitted
given the original control space, so the diminished flexibility is hardly apparent.

\subsection{On the Choice of Reward Functions}
All previous experiments assumed only the most elementary reward functions, the
action-dependent function $V^I$ corresponding to equally-profitable inelastic flows
and the time-dependent function $V^E$ rendering an elastic objective to maximize
the probability of meeting the (hard) deadline. The results illustrated how our
maximum-utility formulation, involving parameter $\lambda^I$ in the multi-component
cost function, sensibly balances between these elementary inelastic and elastic
objectives. Our formulation also permits richer reward functions
and, in turn, features versatility for what specific inelastic and elastic objectives
to balance. The impact of unequal reward rates among inelastic flows
is essentially what was represented in
\FigRef{ModelStructureControl}, the two options for reducing
the control space rendering different per-action attributes. This subsection focuses
on the impact of different elastic reward functions, revisiting the scenario of
\sSecRef{BaselineResults} but introducing the notion of a soft deadline.

\begin{figure}[t!]
\centering
\vspace{0.1in}
\includegraphics[width=0.45\columnwidth]{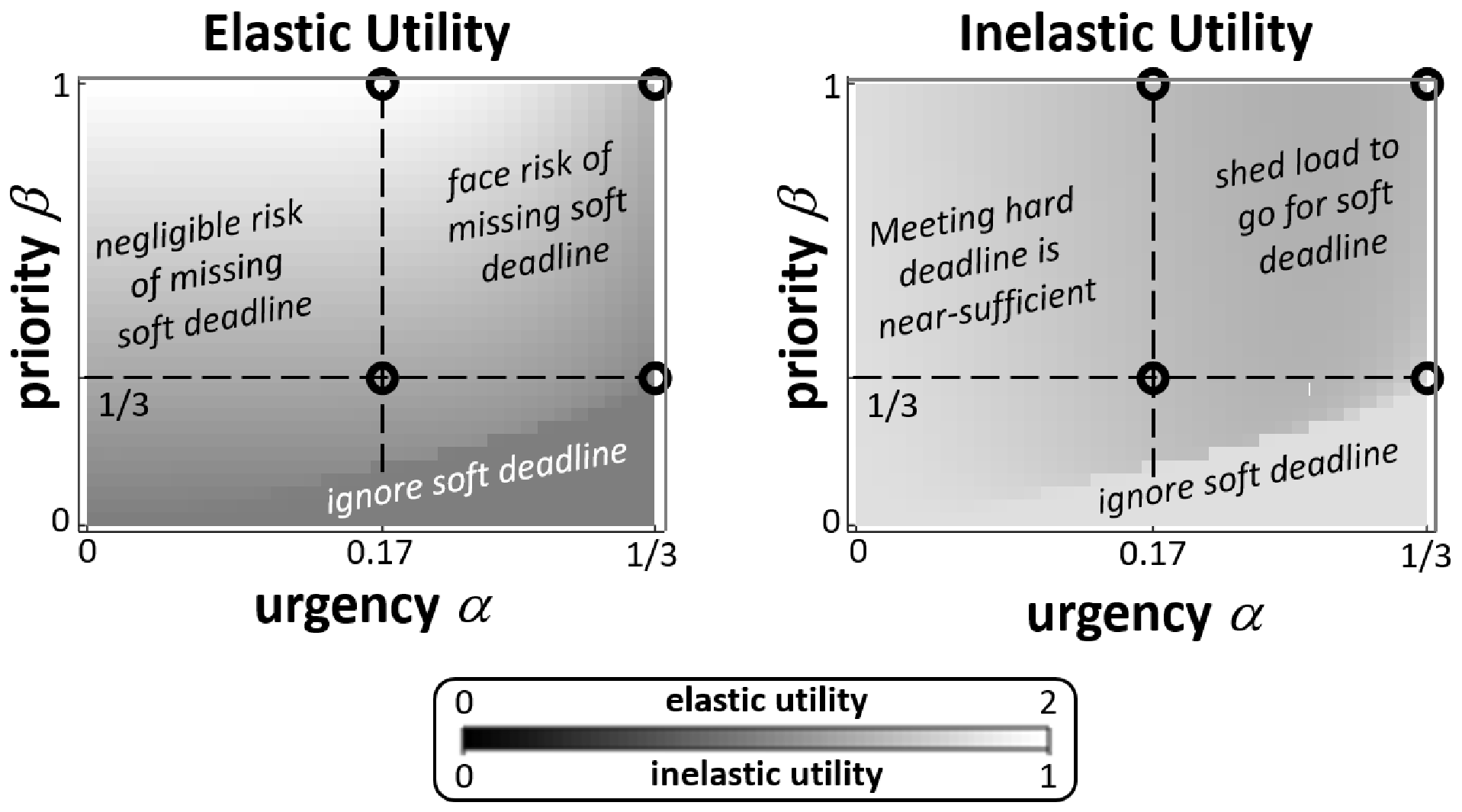} \\
{\footnotesize (a)~Performance (with $\lambda^I = 1$) versus Early Completion Preference} \\[0.2in]
\hspace*{-0.1in}\includegraphics[width=0.475\columnwidth]{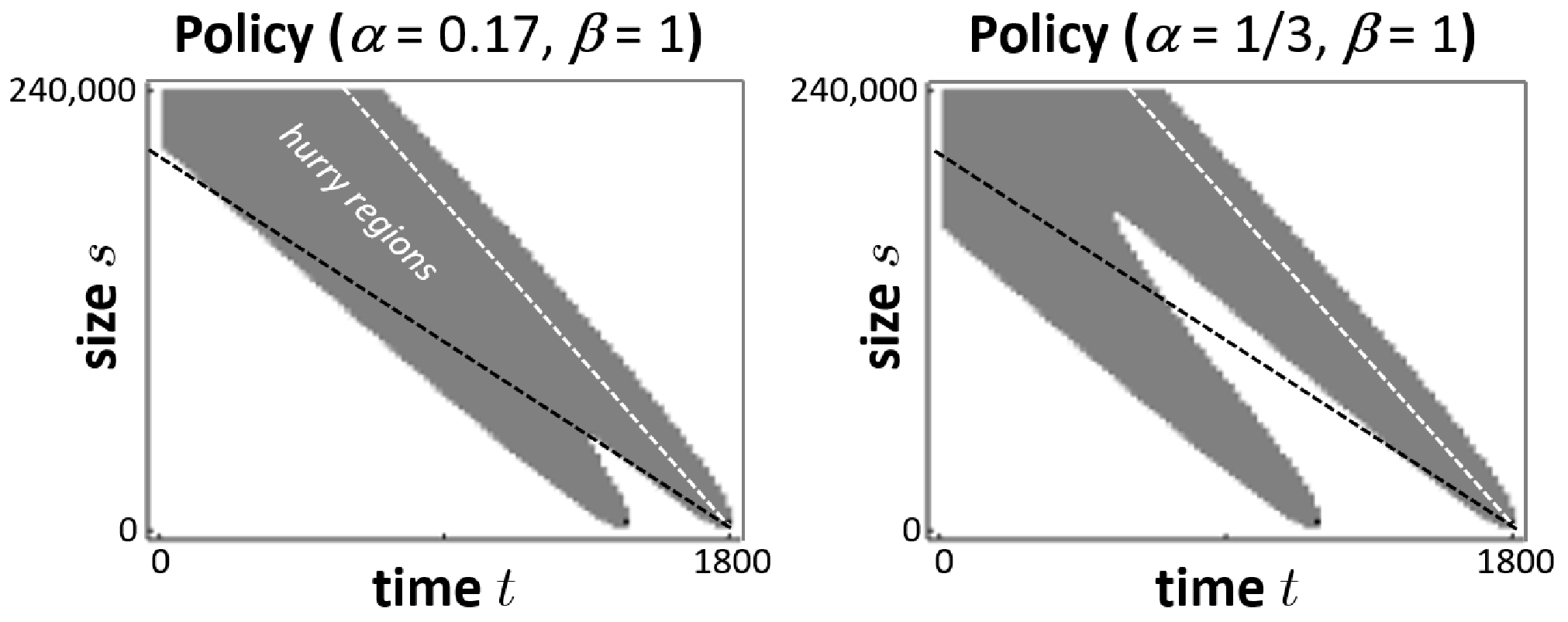} \\[0.1in]
\hspace*{-0.1in}\includegraphics[width=0.475\columnwidth]{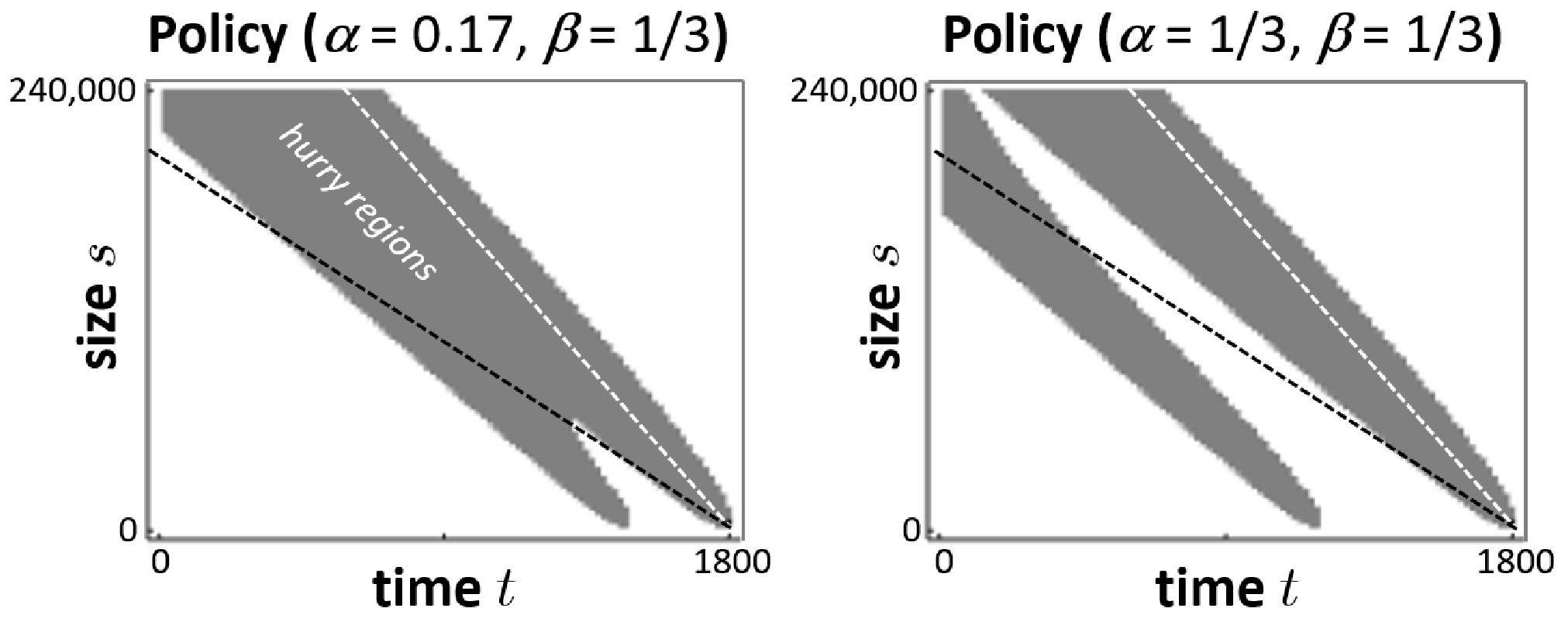} \\
{\footnotesize (b)~Policies at Preferences Indicated by Four Circles in the Utility Plots}
\caption{\label{fig:RewardFunctionsElastic} Results for the
scenario summarized in \FigRef{BaselineScenarioSetup} but with the elastic reward function
expressing a preference for early completion. The urgency of such preference is parameterized by
$\alpha$ in $(0,1/3]$, yielding a soft deadline of time $(1-\alpha)T$, while the priority of
such preference is parameterized by $\beta$ in $(0,1]$, yielding an early completion reward
of $(1+\beta)$ versus the unit-reward of later completion. (a)~The two gray-scale images visualize the
optimized utility as a function of urgency and priority: the elastic
utility fades from dark to light in correspondence with diminishing risk of missing
higher-priority soft deadlines; the inelastic utility fades from dark to light in
correspondence with increasing sustained load. (b)~Four policies, corresponding to
preferences indicated by the four circles in the utility plots, demonstrate that a soft
deadline invites the potential to form a second ``hurry'' region in the state space but the
details of that second hurry region, both its shape and location, are influenced by priority
and urgency.}
\end{figure}

A soft deadline expresses a preference for early completion relative to our hard
deadline $T$. To quantify such preference requires at least two
parameters: one to express the urgency, or the meaning of early, and the other
to express the priority on early completion over later completion. One
way to introduce a soft deadline, starting with any elastic reward function
$\bar{V}^E$ without an early completion preference, is via the modification
$$
V^E(t) = \left\{
\begin{array}{lcl}
(1+\beta)\bar{V}^E(t) & , & 0 < t \leq (1-\alpha)T \\
\bar{V}^E(t) & , & \mbox{otherwise}
\end{array}
\right.
$$
where $\alpha \in (0,1)$ parameterizes the urgency and $\beta>0$ parameterizes the
priority. \FigRef{RewardFunctionsElastic} summarizes the results of applying this
modification to the scenario of \FigRef{BaselineScenarioSetup}, showing both the elastic
and inelastic utilities achieved by the optimized policy as the soft deadline's urgency
and priority varies. Only urgency with
$\alpha\leq 1/3$ is considered because it is on-average infeasible to complete the inelastic
flow earlier than time $S/B$, which in our scenario corresponds to time $2T/3$;
similarly, priority with $\beta > 2$ is not considered as in our scenario the results
are hardly different from those with $\beta = 2$. Preferences with priorities below an urgency-dependent minimum
will simply ignore the soft deadline, reverting to the policies of
\FigRef{BaselineScenarioResults}. Only preferences with relatively high
urgency and high priority will have substantial bearing on the policies, essentially
replicating the ``hurry'' region discussed for hard deadlines but in the portion
of the state space aligned to the soft deadline.

% \textcolor{red}{How do results generalize for inelastic reward functions that are not equally-profitable? Consider a VoIP reward
% rate that is a fraction $\beta < 1$ of the video reward rate, examining results for different choices of $\beta$.}

\section{Augmented Problem Formulations \label{sec:AugmForm}}
The results in the previous section speak only to ideal controller behavior in the sense
that no provisions are made for the possibility of errors in model parameters.
While the stochastic representation tolerates a variance in the model-based
predictions, reliably estimating the link bandwidth or inelastic loads during actual operation
is challenging so the prospect of biased predictions ought not be ignored. \sSecRef{RobAnal}
takes a step in this direction and extends the formulation to include (i)~a possibility of
detrimental congestion, occurring upon admission of inelastic load in excess
of the link's bandwidth, and (ii)~a desired minimum
elastic rate, providing a mechanism by which to tune policies for varying degrees of
robustness to mis-estimated parameters. The original formulation also assumes ``stateless''
inelastic traffic in the sense that profit accrues additively over time no matter the
specific pattern of suspension or readmission---it is only the sum-total duration
in admission that determines inelastic utility. Some inelastic traffic
is truly profitable only under restricted admission patterns e.g., not tolerating
an arbitrary delay before first admission nor interruptions
of arbitrary number and duration after first admission.
\sSecRef{StateAug} discusses state
augmentations to express such richer inelastic objectives, albeit at the expense of
increased problem complexity.

\subsection{On Robustness to Mis-Estimated Model Parameters \label{ssec:RobAnal}}
This subsection augments the formulation for the possibility of
mismatch between the model assumed during policy optimization and the model
presented during policy implementation. We refer to the former as
the nominal model, the associated stochastic shortest path problem denoted by
matrices $\bar{\mat{F}}$ and $\bar{\mat{G}}$ that via optimization yields
policy $\bar{\mu}^*$. With omniscience of the true model, or matrices $\mat{F}$ and
$\mat{G}$ that via optimization yields the policy $\mu^*$, an additional policy
evaluation (of nominal policy $\bar{\mu}^*$ against the true matrices) permits
the impact of model mismatch to be quantified by the difference of expected
sum-total cost vectors, $\mat{J}(\bar{\mu}^*) - \mat{J}(\mu^*)$. This
difference between nominal and omniscient performance is separable into
its elastic and inelastic components, also evaluating both policies against
matrices $\mat{F}$ and $\mat{G}^E$ as was discussed in \sSecRef{TheAlgs}.
Note that this procedure applies only for mismatch that preserves model
structure, which fortunately includes the primary cases of practical
interest e.g., mis-estimated bandwidth/load parameters or otherwise
erroneous risk predictions. It remains to generalize the probability and cost
calculations so that, no matter the mismatch, both nominal and true matrices
are well-defined.

\subsubsection{Congestion and Desired Minimum Elastic Rates}
Detrimental congestion is modeled by the circumstance in which admitted
inelastic load exceeds the link's bandwidth, which we assume not only starves
the inelastic flow but also prohibits the usual inelastic reward.
Such control would never be purposefully invoked, so assuming
that every action's mean elastic rate $\bar{R}(a)$ is non-negative suffices
for the nominal model. With model mismatch, however, it
is possible that the (true) inelastic load $L(a)$ exceeds the (true)
bandwidth $B$. We generalize the per-action
calculations of mean elastic rate and inelastic reward rate, respectively, to
$$
R(a) = \max\{0,B-L(a)\}
$$
and
$$
V^I(a) = \left\{
\begin{array}{lcl}
\bar{V}^I(a) & , & \mbox{if $L(a) \leq B$} \\
0 & , & \mbox{if $L(a) > B$}
\end{array}
\right. ,
$$
where $\bar{V}^I$ denotes the (calibrated) inelastic reward function in
the nominal model. It is also worth noting that, in the true model, the
inequalities of \equRef{ActionSpace} or \equRef{InelasticRewards} are no
longer guaranteed.

The notion of a desired minimum elastic rate, which we will denote by $R^0$, invites
numerous interpretations. In the strictest sense, it poses a hard constraint by which to
reduce the control space to only those actions for which the (nominal) inelastic rate $\bar{R}(a)$ exceeds
$R_0$. In the loosest sense, it could be expressed as a soft deadline
with urgency $\alpha = 1 - S/(R^0T)$. An interpretation between these two extremes,
in essence posing a soft rate constraint, offers the most direct mechanism to tune our policies
for varying degrees of robustness. Specifically, let
function
$$
C^0(s,t) = \left\{
\begin{array}{lcl}
-V^E(T)/T & , & \mbox{if $s > S-R^0t$} \\
0 & , & \mbox{if $s \leq S-R^0t$}
\end{array}
\right.
$$
assign a high cost rate (in negated utils per second) whenever the
elastic status falls behind the predicted mean progress under desired
rate $R^0$;  this cost rate is high in the sense of being equal in magnitude to the
(nominal and calibrated) inelastic reward rate of action $m$. We then
generalize \equRef{DualObjective} to involve three components
$$
\mat{G}(a) = \mat{G}^E(a) + \lambda^I\mat{G}^I(a) + \mat{G}^0(a),
$$
the third matrix populated for every action $a$ as follows: to every
probable transition from state $i \leftrightarrow (x,k)$ to state $j \leftrightarrow (x^\prime,k+1)$
such that step $x < M$ and stage $k < N$, assign cost $G^0_{i,j}(a) = C^0(s,t)\delta_T$ with
size $s = S - x^\prime\delta_S$ and time $t = (k+1)\delta_T$. It is worth noting that specifying
$R^0 = 0$ reverts cost matrix $\mat{G}$ to its original formulation.

\subsubsection{Examples and Experiments}
We now revisit the scenario of \sSecRef{BaselineResults} with the generalizations just discussed. Specifically,
we allow for the possibility of congestion and examine the use of a minimum desired elastic rate $R^0$ to tune between
performance (under nominal conditions) and robustness (under mismatched conditions). For brevity, the
experiments consider only mismatch in the form of an over-estimated link bandwidth, fixing parameter
$\bar{B} = 200$ in the nominal model but varying the true bandwidth $B$ to represent different degrees of
over-estimated resources. The situations are analogous to cases of under-estimated inelastic loads or
otherwise overly-optimistic risk predictions.

\begin{figure}[t!]
\centering
\vspace{0.1in}
\includegraphics[width=0.45\columnwidth]{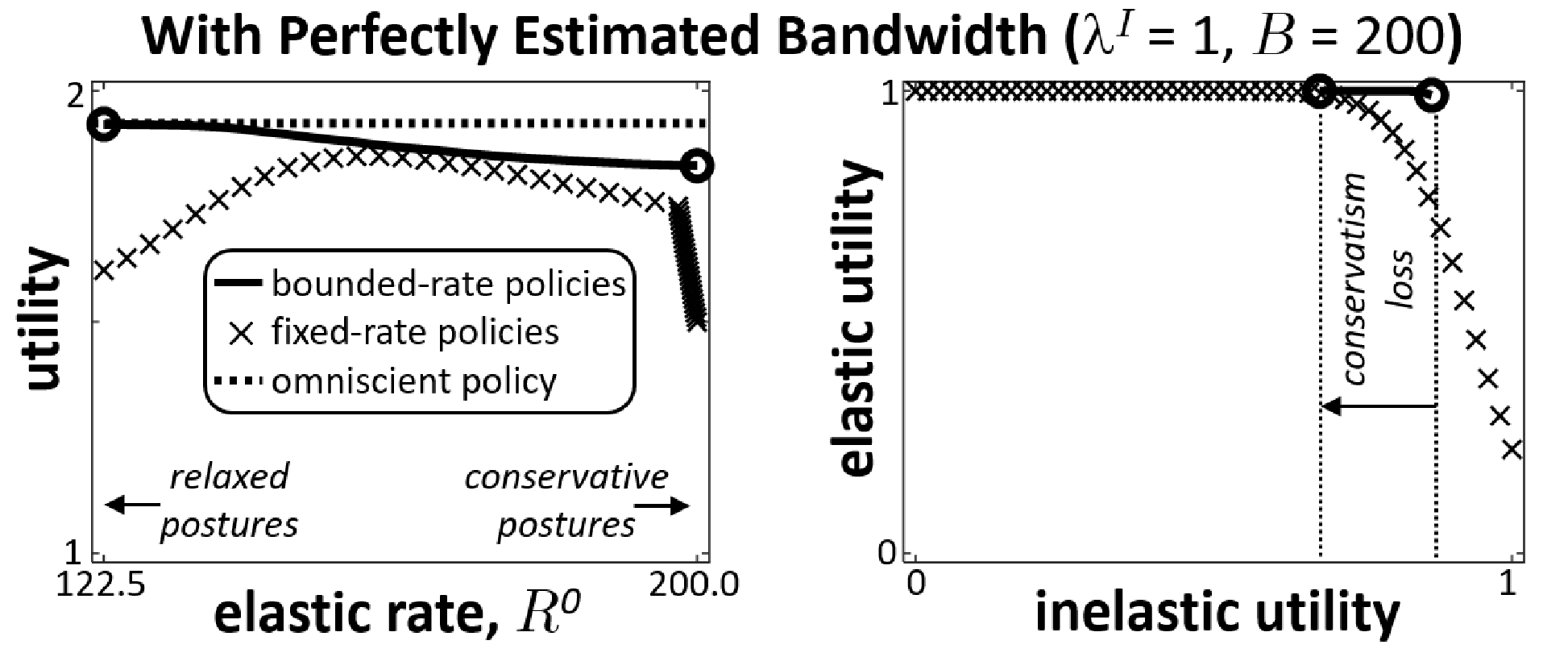} \\[0.1in]
\hspace*{-0.1in}\includegraphics[width=0.475\columnwidth]{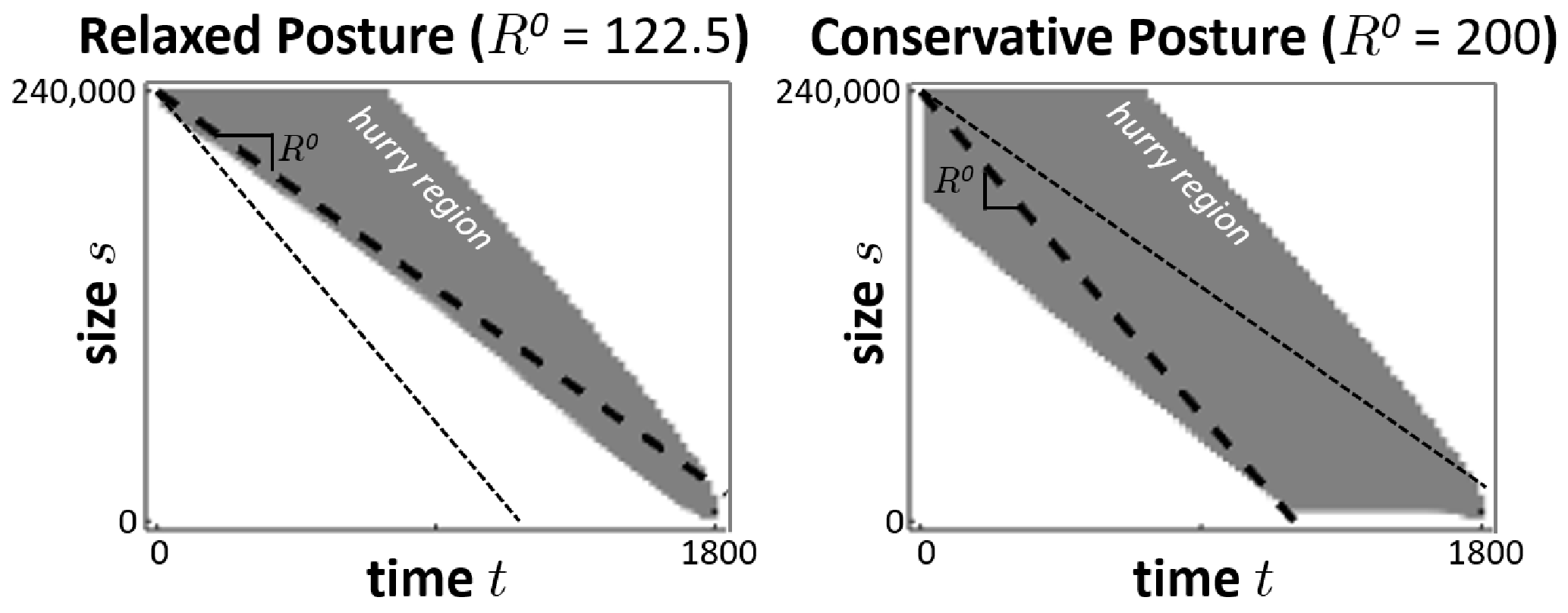} \\
{\footnotesize (a)~Performance Loss of Rate-Bounded Policies: Behavioral Details} \\[0.2in]
\includegraphics[width=0.45\columnwidth]{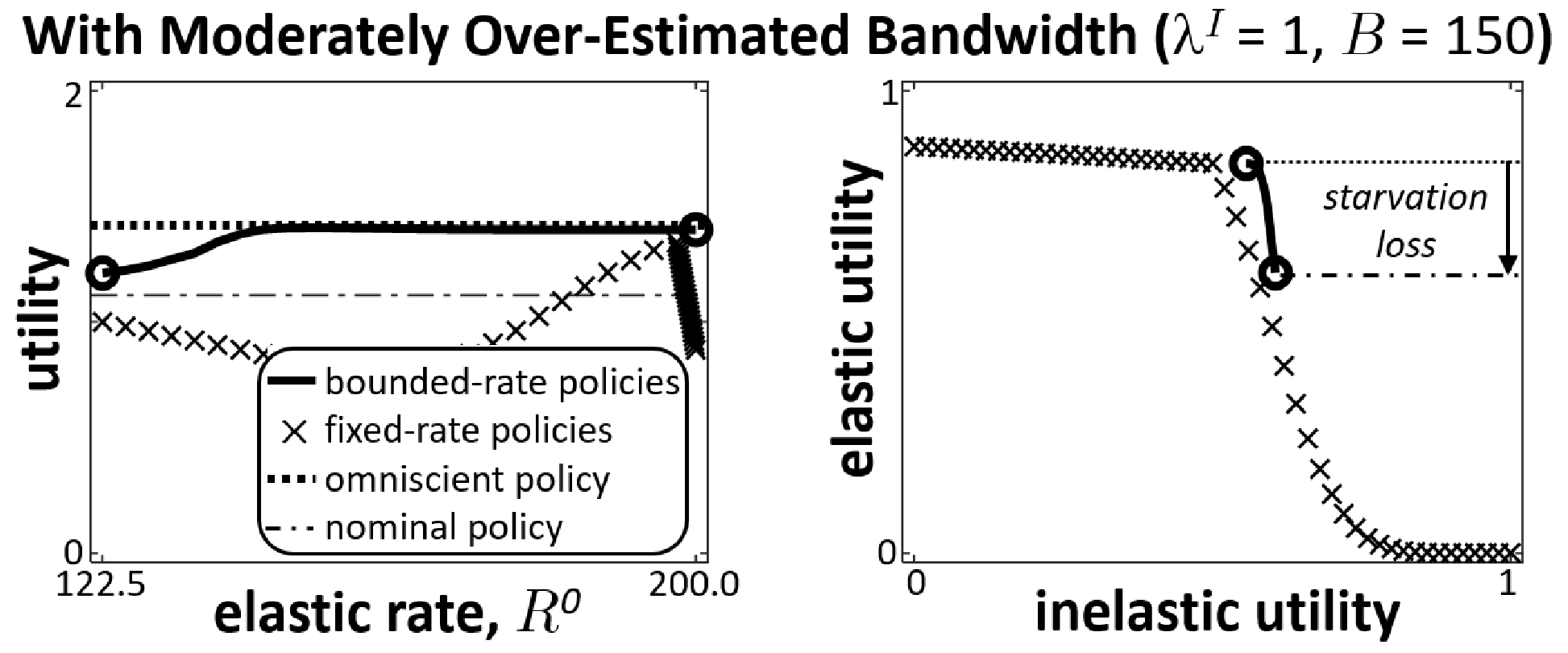} \\[0.1in]
\includegraphics[width=0.45\columnwidth]{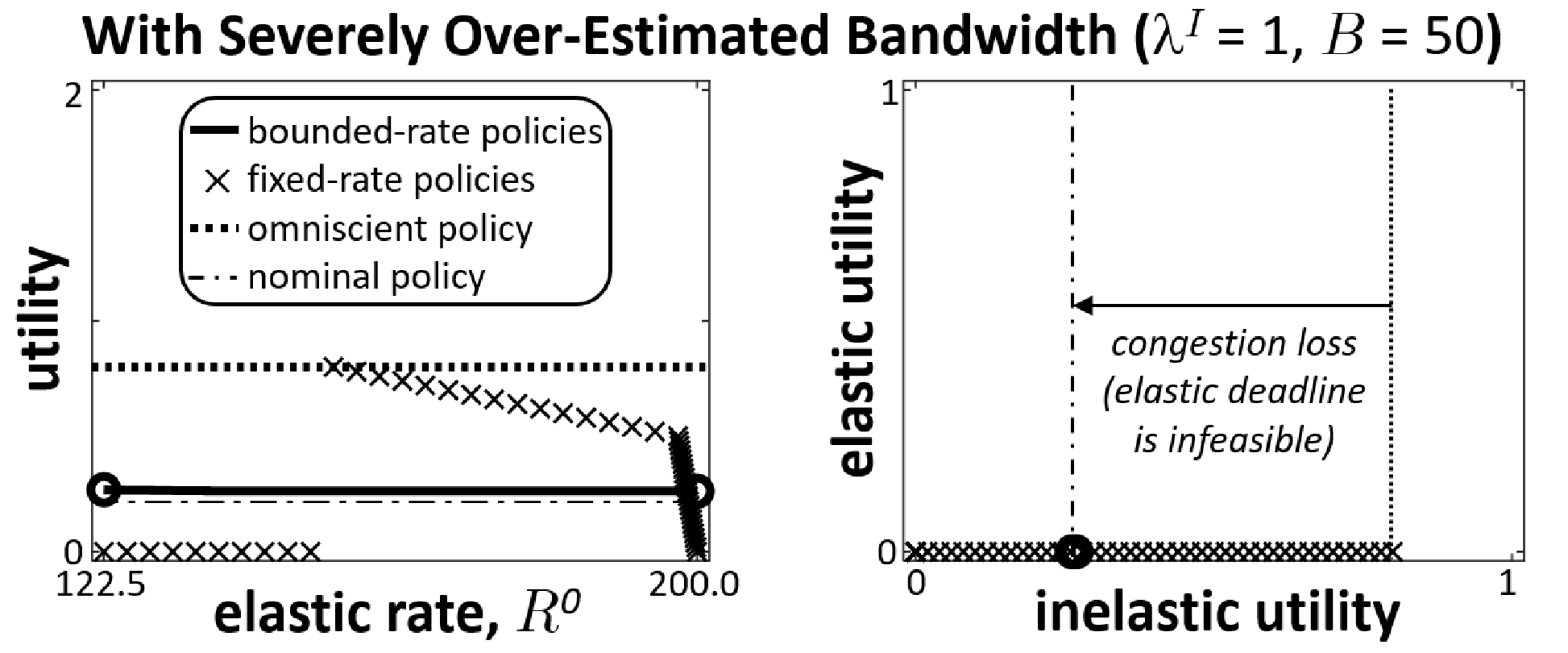} \\[0.1in]
\hspace*{-0.1in}\includegraphics[width=0.475\columnwidth]{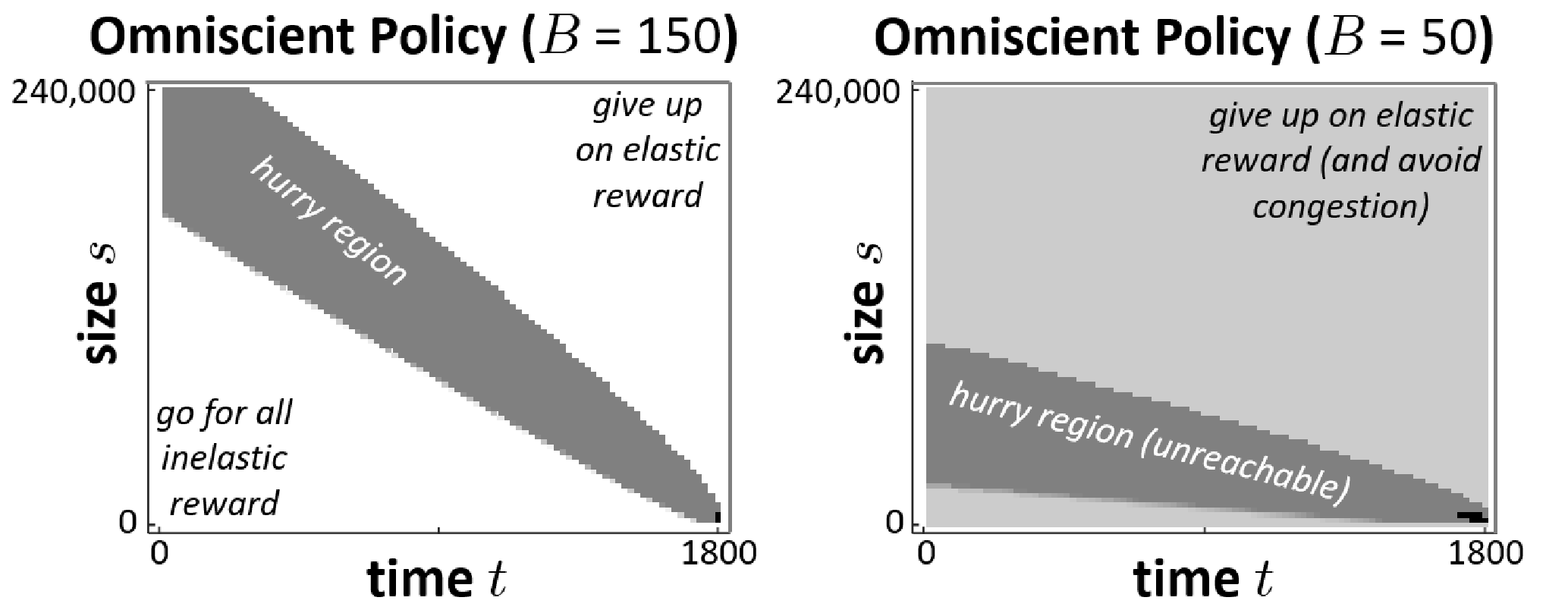} \\
{\footnotesize (b)~Robustness Gain of Rate-Bounded Policies: Behavioral Details}
\caption{\label{fig:BaselineScenarioRobustness} Results of the performance-robustness analysis
to over-estimated link bandwidth $B$ in the scenario summarized in \FigRef{BaselineScenarioSetup} upon
augmenting the formulation to include (i)~the possibility of detrimental congestion and
(ii)~the option to specify a desired minimum elastic rate $R^0$. The omniscient policy given $B = 150$
in (b) has attributes exhibited by the bounded-rate policies in (a), so tuning
$R^0$ can balance between the conservatism loss identified in (a) and the
starvation loss identified in (b). Comparable robustness to the congestion loss also identified in (b) is impossible as no
choice of rate bound $R^0$ will mimic the omniscient policy given $B = 50$.}
\end{figure}

\FigRef{BaselineScenarioRobustness}(a) summarizes
the results of the bounded-rate policies under nominal conditions (with $B = \bar{B}$). As the bound
$R^0$ increases, the controlled behavior is swept from ``relaxed postures'' that like the nominal policy
hurry the elastic flow primarily as the deadline nears (and admit all inelastic flows early) to
``conservative postures'' that hurry the elastic flow early (and admit all inelastic flows upon completion).
Under nominal conditions, only the conservative postures result in diminished inelastic utility,
identified as conservatism loss, and the bounded-rate policies consistently balance between
elastic and inelastic objectives better than their fixed-rate counterparts.
\FigRef{BaselineScenarioRobustness}(b) provides results for two cases of model mismatch,
demonstrating the robustness gain that is achievable by the bounded-rate policies. In the case of only moderate
model mismatch (e.g., with $B = 150$, or a 33\% over-estimate of bandwidth), the bounded-rate policies are
seen to maintain near-omniscient performance for roughly the same postures that under nominal conditions exhibited
conservatism loss---moreover, only the most-relaxed postures are comparably susceptible to the
nominal policy's unintentional starvation of the elastic flow. The degree of model mismatch
for which such robustness gain is achievable is limited, of course: our selected case of severe model
mismatch (e.g., with $B = 50$, or 300\% over-estimate of bandwidth) demonstrates that such
limits are entwined with the advent of congestion. \FigRef{BaselineScenarioRobustness} also compares
four policies, the most-relaxed and most conservative bounded-rate ones in (a) with
the two omniscient ones in (b), offering further insights into the controlled behaviors underlying the
performance-robustness tradeoff and their dependence on scenario details. The omniscient policy given $B = 150$
has attributes exhibited by the bounded-rate policies through suitable choice of $R^0$, which is
not the case given $B = 50$.

\subsection{On Richer Inelastic Objectives \label{ssec:StateAug}}
This subsection augments the formulation of \SecRef{ProbForm} for richer inelastic objectives
than simply maximizing the expected sum-total duration in admission. Recall in \FigRef{BaselineScenarioResults}(a)
the fairly rapid oscillation of the control signal during the time period in which the elastic
flow is hurried. What if a subset of the video flows are not well-served by such rapid suspension
and readmission, rather requiring patterns with interruptions that are
limited in number or duration? Such required persistence implies that, to truly realize the
associated inelastic reward, the decision to admit or readmit these video flows must more
judiciously account for the commitment of future resources than is the case for their
stateless counterparts. In turn, depending on the commitment's increased risk to missing
the elastic deadline, a control trajectory in which the video flows are suspended throughout
the hurry period of \FigRef{BaselineScenarioResults}(a) may net greater total utility.
On the other hand, what if there is also an urgency requirement, in the sense that
the profitable patterns also cannot tolerate arbitrary delay of (re)admission?

\subsubsection{Inelastic Flows Requiring Persistence or Urgency}
\begin{figure}[t!]
\centering
\vspace{0.1in}
\includegraphics[width=0.475\columnwidth]{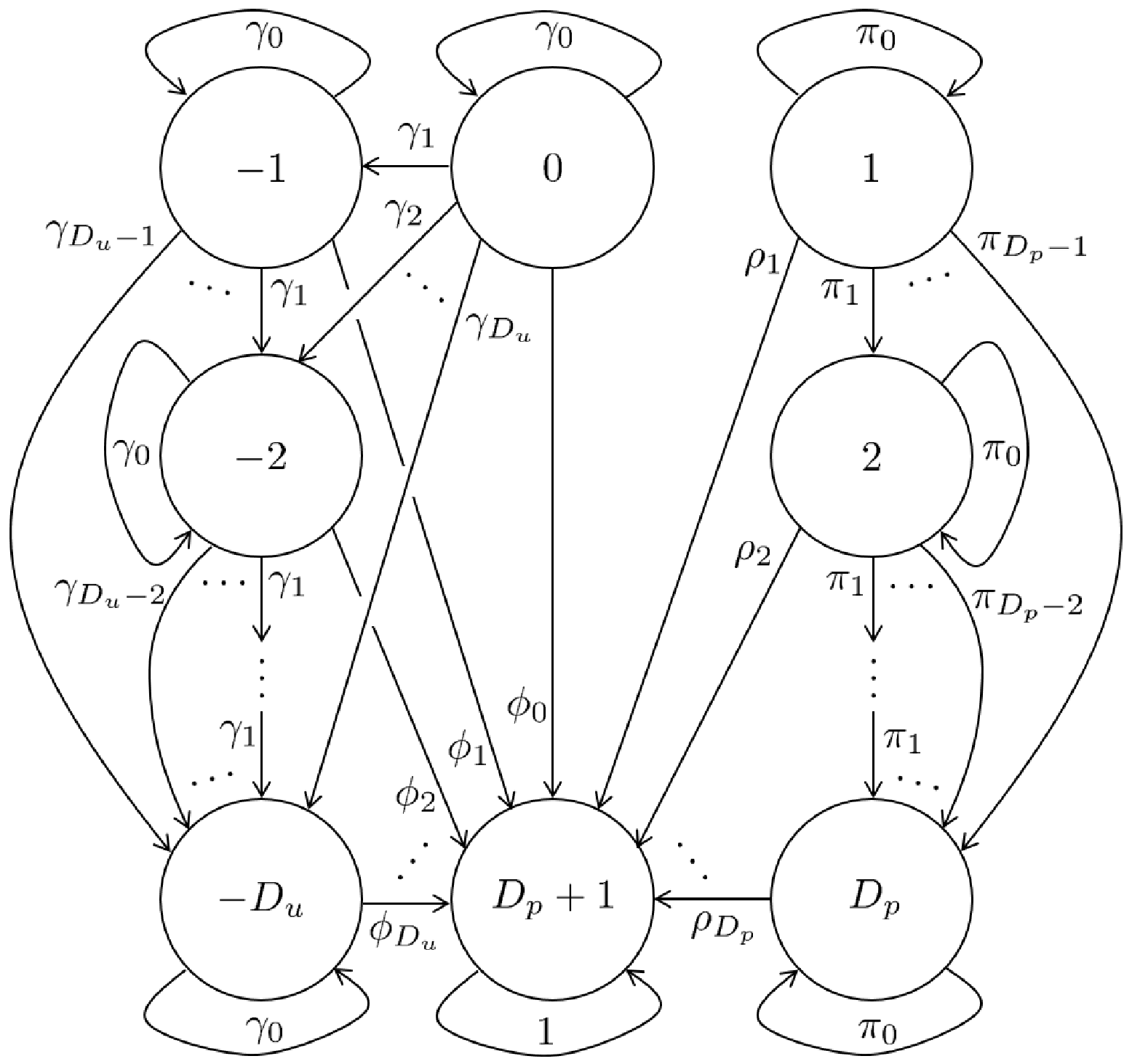} \\
(a)~Under Actions that Deny the Stateful Inelastic Traffic \\[0.2in]
\includegraphics[width=0.475\columnwidth]{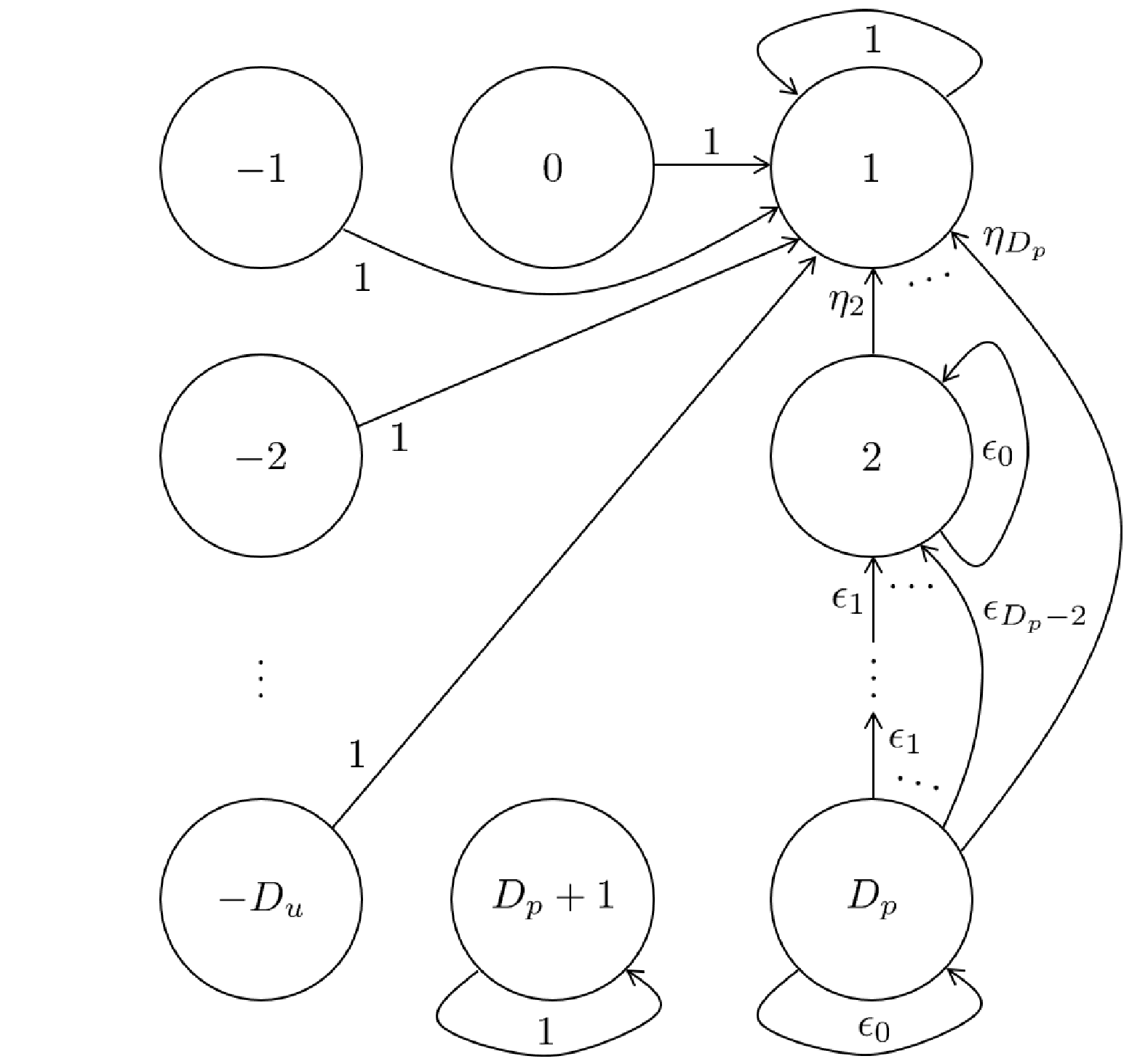} \\
(b)~Under Actions that Admit the Stateful Inelastic Traffic
\caption{\label{fig:InelasticStates} Probable single-stage
state transitions among ${D = D_p + D_u + 2}$ levels for
representing inelastic objectives with persistence or urgency requirements.
Level~0 is the initial state and urgency-related levels
$-1, -2, \ldots, -D_u$ represent states of diminishing opportunity for profitable first-time
admission. Level~$1$ represents the freshest state of
profitable (re)admission and persistence-related levels $2, 3, \ldots, D_p$
represent states of diminishing freshness. Level $D_p+1$ represents
a state of permanent suspension because (re)admission can no longer
be profitable, which is entered from positive (non-positive) levels only upon
failing to meet persistence (urgency) requirements. Actions that (a)~deny
the stateful inelastic traffic render dynamics that progress towards
permanent suspension (level $D_p+1$), while (b)~otherwise the progress is towards
freshest admission (level $1$). The single-stage transition
probabilities, or parameters $\gamma_{-w}$ and $\phi_{-w}$ for urgency-related
levels $w\leq 0$ as well as parameters $\pi_{w-1}$, $\rho_w$, $\epsilon_{w-1}$ and $\eta_w$
for persistence-related levels $w \geq 1$, are derived from given stochastic counting
processes (in levels per second) that are independent of those analogously
employed within the elastic flow's risk predictions.}
\end{figure}

Requirements of persistence or urgency are represented in our
formulation by generalizing the model structure discussed in \sSecRef{TheStruc},
both state space and control space. The generalized state space introduces
two integers $D_p\geq 1$ and $D_u \geq 0$ that determine $D = D_p + D_u + 2$ discrete
inelastic states, or levels, indexed by
$$
w = -D_u, -D_u+1, \ldots, -1, 0, 1,
\ldots, D_p, D_p+1
$$
as illustrated in \FigRef{InelasticStates}.
The number of (global) states increases to ${n = D(M+1)(N+1)}$,
replicating the elastic flow's size-time discretization in \FigRef{StateSpace}
per level, so the state indices $1, 2, \ldots, n$ are now in correspondence
with level-step-stage triples, $i \leftrightarrow (w,x,k)$. The control space is generalized
to maintain separation of stateless and stateful inelastic traffic. Specifically, if there are $\bar{m}$
admissible subsets of stateless inelastic flows, then the separate choice
to deny or admit a set of stateful inelastic flows yields a composite control space with
$m = 2\bar{m}$ actions. Each composite action $a = (a_1,a_2)$ implies a mean elastic load $L(a)$ and, in turn,
the composite control space can be ordered to satisfy the inequalities of \equRef{ActionSpace}
and \equRef{InelasticRewards} as in the original formulation.

\FigRef{InelasticStates} also illustrates that, in addition to the
structural parameters $D_p$ and $D_u$, there are probability
parameters by which system dynamics can be further configured
for a specific inelastic requirement. The persistence-related
dynamics assume that each stage under
suspension (action ${a_1 = 1}$) will certainly never increase
freshness but the decrease in freshness is modeled stochastically
by a given counting process % with mean rate of $\lambda_p(1)$
(in levels per second), implying the probabilities $\pi_\ell$
for $\ell = 0, 1, 2, \ldots$ of progressing $\ell$ levels within
duration $\delta_T$ of any one stage; each stage under
admission (action ${a_1 = 2}$) has similarly modeled dynamics,
except that only non-decreasing freshness is probable with
another (and possibly different) counting process implying
probabilities $\epsilon_\ell$. Proceeding analogously to
the risk predictions in \sSecRef{TheProbs}, defining
$\rho_w = \sum_{\ell = D_p+1-w}^\infty \pi_\ell$
and
$\eta_w = \sum_{\ell = w-1}^\infty \epsilon_\ell$
for levels $w = 1, 2, \ldots, D_p$, the persistence-related
transition probabilities express a pair of {$(D_p+1)$-by-$(D_p+1)$} stochastic matrices,
one per inelastic action $a_1$:
$$
\mat{P}^p(1) =
\left[
\begin{array}{ccccc}
\pi_0    & \pi_1  & \cdots & \pi_{D_p-1} & \rho_1       \\
0        & \pi_0  & \ddots & \vdots      & \vdots       \\
\vdots   & \ddots & \ddots & \pi_1       & \rho_{D_p-1} \\
0        & \cdots & 0      & \pi_0       & \rho_{D_p} \\
0        & 0      & \cdots & 0           & 1\end{array}
\right]
$$
and
$$
\mat{P}^p(2) =
\left[
\begin{array}{cccccc}
1            & 0                & 0          & \cdots     & 0          & 0      \\
\eta_2       & \epsilon_0       & 0          & \ddots     & \vdots     & \vdots \\
\eta_3       & \epsilon_1       & \ddots     & \ddots     & 0          & 0      \\
\vdots       & \vdots           & \ddots     & \epsilon_0 & 0          & 0      \\
\eta_{D_p}   & \epsilon_{D_p-2} & \cdots     & \epsilon_1 & \epsilon_0 & 0      \\
0            & 0                & \cdots     & 0          & 0          & 1
\end{array}
\right] .
$$
The urgency-related dynamics are similarly expressed but, because of the deterministic
transition into persistent states upon first admission, involve only
one additional counting process and probabilities $\gamma_{\ell}$
and $\phi_{-w} = \sum_{\ell=D_u+w+1}^\infty \gamma_\ell$ for
levels $w = -D_u, \ldots, -1, 0$. The pair of
stochastic matrices are each of dimension $(D_u+1)$-by-$D$ to
explicitly express this controllable transition to persistence:
$$
\mat{P}^u(1) =
\left[
\begin{array}{cccc|ccccc}
\gamma_0     & 0        & \cdots   & 0        & 0        & 0      & \cdots & 0           & \phi_{D_u}   \\
\gamma_1     & \gamma_0 & \ddots   & \vdots   & 0        & 0      & \ddots & \vdots      & \vdots       \\
\vdots       & \ddots   & \ddots   & 0        & \vdots   & \vdots & \ddots & 0           & \phi_1       \\
\gamma_{D_u} & \cdots   & \gamma_1 & \gamma_0 & 0        & 0      & \cdots & 0           & \phi_0
\end{array}
\right]
$$
and
$$
\mat{P}^u(2) =
\left[
\begin{array}{cccc|ccccc}
0      & 0      & \cdots   & 0      & 1            & 0                & \cdots & 0          & 0      \\
0      & 0      & \ddots   & \vdots & 1            & 0                & \ddots & \vdots     & \vdots \\
\vdots & \ddots & \ddots   & 0      & \vdots       & \vdots           & \ddots & 0          & 0      \\
0      & \cdots & 0        & 0      & 1            & 0                & \cdots & 0          & 0
\end{array}
\right] .
$$
Altogether, the following $D$-by-$D$ stochastic matrices fully
characterize our stateful inelastic traffic:
$$
\mat{P}^I\!\!\left( a_1\right) = \left[
\begin{array}{c|c}
\multicolumn{2}{c}{\mat{P}^u\!\left( a_1\right)} \\[0.1in]
\mat{0} & \mat{P}^p\!\left( a_1 \right)
\end{array}
\right] , \quad a_1 \in \{ 1, 2 \} \quad .
$$

Having (i)~augmented the state space, originally the grid of step-stage pairs $(x,k)$, with the size-$D$ third dimension over levels $w$,
(ii)~structured the control space to treat stateful and stateless inelastic traffic separately via composite actions $a = (a_1,a_2)$ and (iii)~characterized
the system dynamics over levels with action-dependent $D$-by-$D$ stochastic matrices $\mat{P}^I(a_1)$, it remains to capture these generalizations
in a stochastic shortest path problem defined by per-action matrices $\mat{F}(a)$ and $\mat{G}(a)$. All aspects discussed for the original formulation
remain pertinent so let us start from the per-action matrices assuming only stateless inelastic flows, which we denote by $\tilde{\mat{F}}(a)$, $\tilde{\mat{G}}^E(a)$ and $\tilde{\mat{G}}^I(a)$. We assume that the counting processes underlying both the elastic risk predictions and the inelastic level dynamics are mutually independent. It follows for the augmented model that
every transition from state $i \leftrightarrow (w,x,k)$ to state ${j \leftrightarrow (w^\prime, x^\prime,k+1)}$ under action $a$ is assigned the probability
$$
F_{i,j}(a) = \tilde{F}_{\tilde{i},\tilde{j}}(a)P^I_{w,w^\prime}(a_1),
$$
where $\tilde{i} \leftrightarrow (x,k)$ and $\tilde{j} \leftrightarrow (x^\prime,k+1)$ denote the appropriate indices in the original state space.
Moreover, each such transition bears the original cost except when admitting stateful inelastic traffic in permanent suspension i.e.,
assign elastic cost $G^E_{i,j}(a) = \tilde{G}^E_{\tilde{i},\tilde{j}}(a)$ without exception and inelastic cost
$G^I_{i,j}(a) = \tilde{G}^I_{\tilde{i},\tilde{j}}(a)$ unless $w = D_p+1$ and $a_1 = 2$ in which case
$$
G^I_{i,j}(a) =
\tilde{G}^I_{\tilde{i},\tilde{j}}(a) + \delta_T\left[ V^I\!\left( (2,a_2)\right) - V^I\!\left( (1,a_2) \right)\right],
$$
reducing the instantaneous inelastic reward rate to only that associated with the admitted stateless flow(s).
%Observe that upon entering permanent suspension the admission of stateful inelastic flows will never be
%purposefully invoked because the reduced elastic rate renders no gain in inelastic reward.
%netting the same inelastic reward
%no additional inelastic reward for the additional load over the

Our augmented formulation for richer inelastic objectives merits some reflection. Firstly, a spectrum of urgency and persistence requirements are possible through choices of integers $D_p$ and $D_u$ as well as the various transition probabilities---let us highlight some special cases. With respect to persistence upon first admission, choosing $D_p = 1$ with probability $\pi_1 = 1$ will strive for no disruption, while choosing $D_p$ relatively large with the same counting process under each action will strive for suspension and readmission at 50\% duty cycle on-average---the period of each such cycle is determined primarily by the self-transition probabilities $\pi_0 = \epsilon_0$, while different percentages on-duty can be achieved by choosing counting processes with action-dependent mean arrival rates. Urgency can be entirely ignored by choosing $D_u = 0$ and $\gamma_0 = 1$, while choosing $\gamma_1 = 1$ will permit exactly $D_u$ stages of delay before a profitable first admission is a foregone opportunity---all other things equal, increasing $\gamma_0$ or increasing $D_u$ relaxes the urgency requirement.
Secondly, our augmented formulation was presented assuming only a single set of stateful inelastic flows. The augmentation can in
principle be repeated given multiple such sets, each presumably with a different persistence or urgency
requirement, but the model structure's increasing dimensions quickly challenges scalability. Scalability of the control space can be addressed
in some cases by scenario-dependent considerations as exemplified by \FigRef{ModelStructureControl}, preserving controller fidelity with a number of composite actions not necessarily larger than its counterpart when persistence and urgency are ignored. Scalability of the state space, however, can only be addressed by a coarsened discretization, sacrificing controller agility as exemplified by \FigRef{ModelStructureState}.

\subsubsection{Examples and Experiments}
\begin{figure}[h!]
\centering
\vspace{0.1in}
\includegraphics[width=0.475\columnwidth]{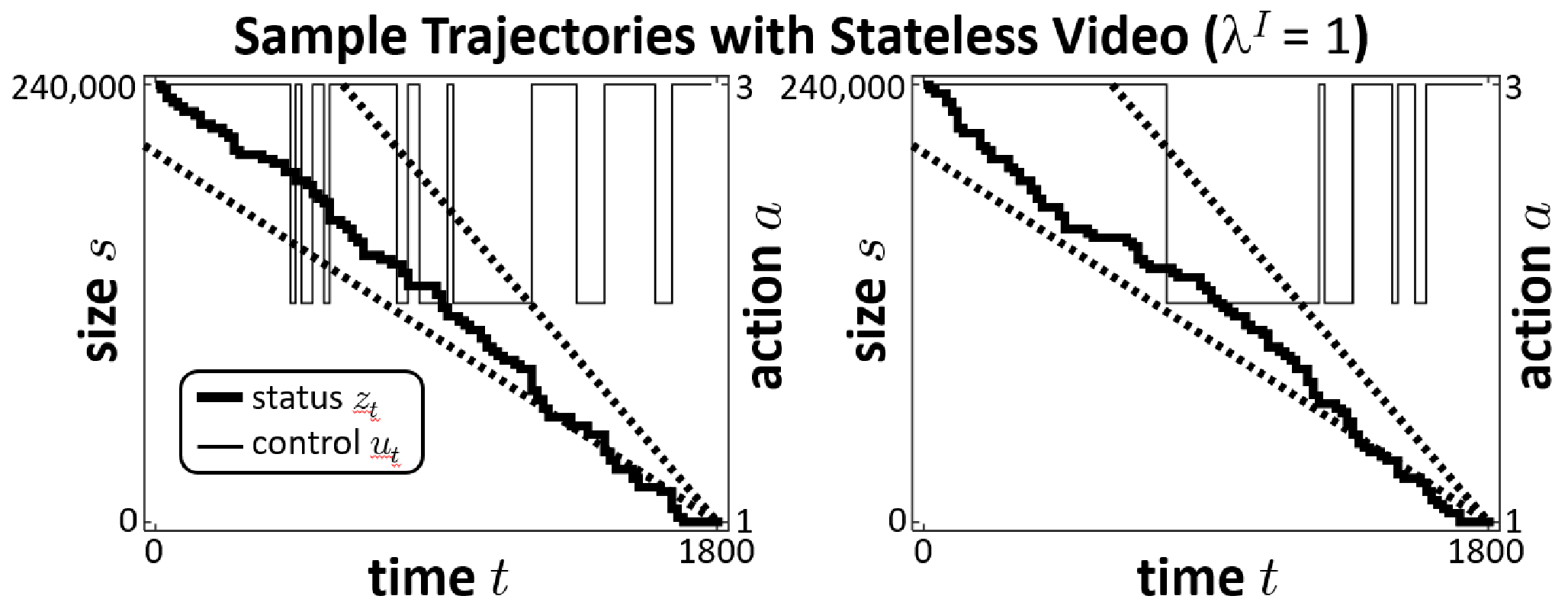} \\
\caption{\label{fig:StatelessInelasticResults} Two sample trajectories of the scenario
summarized in \FigRef{BaselineScenarioSetup} upon reducing the control space to only the
$m = 3$ dominant actions: action~1 denies all 50 inelastic flows, action~2 admits only the
25 VoIP flows and action~3 admits the remaining 25 video flows. Rapid oscillation
of the control signal or just lengthy suspensions before a readmission may not truly
render the vide as profitable. In the case of only stateless inelastic
flows, all 50 are initially admitted (action~3) and the video flows will be
suspended (action~2) for a sufficient sum-total duration to meet the elastic deadline but not
necessarily in preferable patterns.}
\end{figure}

We now revisit the scenario of \sSecRef{BaselineResults} but with richer inelastic objectives on all video flows,
while all VoIP flows remain stateless. Recall that with only stateless inelastic flows,
the optimized policies  (given $\lambda^I = 1$) continuously admit all VoIP flows, capturing half of the inelastic profit
with relatively little load---all video flows are admitted initially but typically suspended then readmitted
thereafter to control for elastic risk. Not all patterns of suspension and readmission may be equally desirable for
the video application at hand, as presented in \FigRef{StatelessInelasticResults}. Our final set of experiments will
reduce the control space to just the $m = 3$ dominant actions, namely to admit (1)~no inelastic flows, (2)~only the VoIP flows or
(3)~both all VoIP and video flows by employing the construction scheme described in \equRef{aggregation} with
$$
\bar{L}_\ell = \left\{
\begin{array}{lcl}
2.5 & , & \ell = 1 \\
75 & , & \ell = 2
\end{array}
\right. \quad \mbox{and} \quad
\bar{V}_\ell = \left\{
\begin{array}{lcl}
25 & , & \ell = 1 \\
25 & , & \ell = 2
\end{array} \right. .
$$
Upon calibration with elastic reward $V^E(T) = 1$, the inelastic
reward function becomes $V^I(a) = 0.5(a-1)/T$.
Despite the reduced flexibility, for reasons exemplified by
\FigRef{ModelStructureControl}, the admission controller
is near-equivalent to that in \FigRef{BaselineScenarioResults}(a)
over the original per-flow control space. It's also worth noting
for this scenario that admitting only the video
flows is a superfluous fourth action, yielding no greater inelastic reward
rate than only the VoIP flows of lesser load.

\begin{figure}[t!]
\centering
\vspace{0.1in}
\includegraphics[width=0.475\columnwidth]{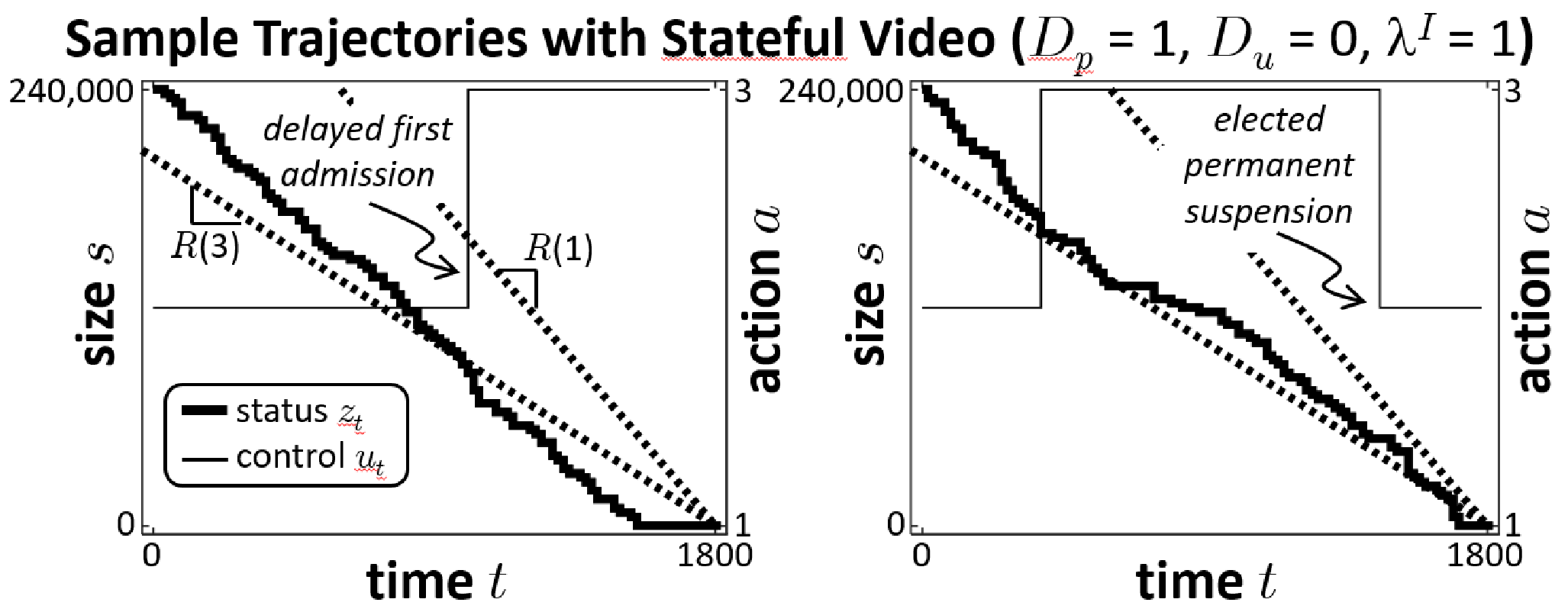} \\[0.1in]
\includegraphics[width=0.475\columnwidth]{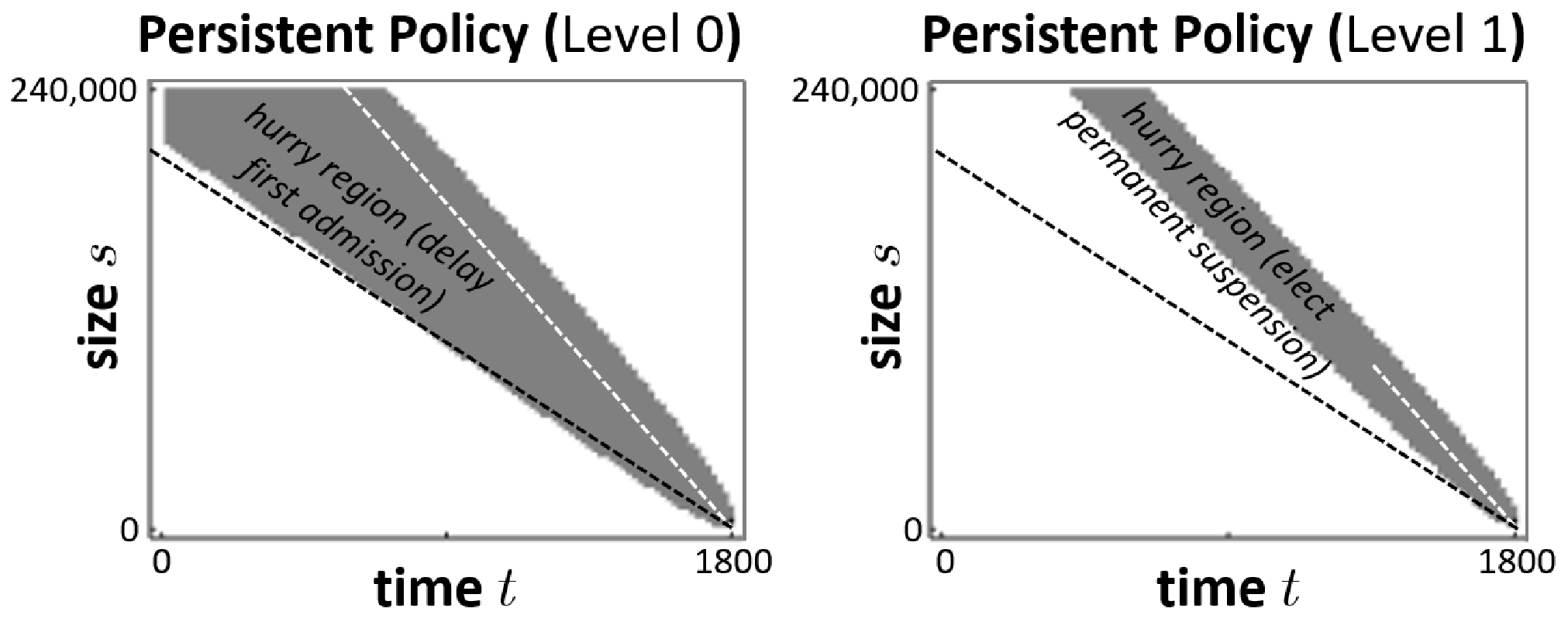} \\
(a)~With Inelastic Requirements of Persistence Only\\[0.2in]
\includegraphics[width=0.475\columnwidth]{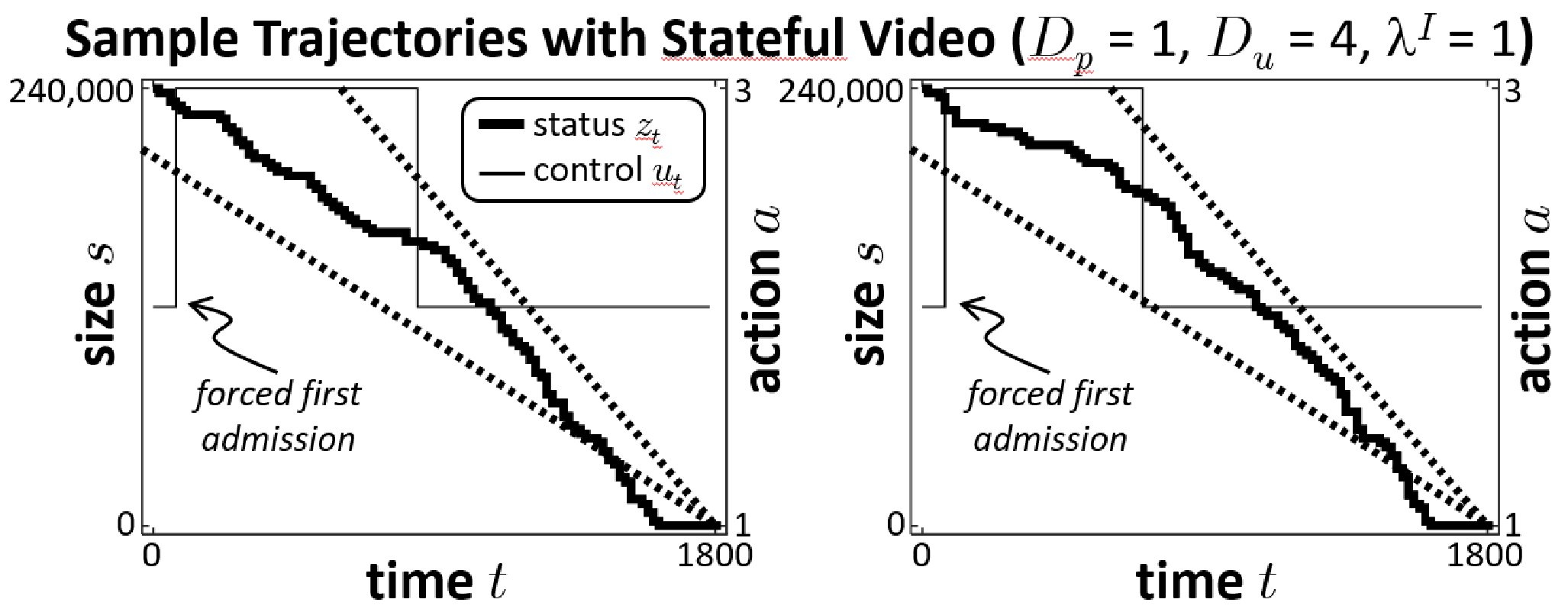} \\[0.1in]
\includegraphics[width=0.475\columnwidth]{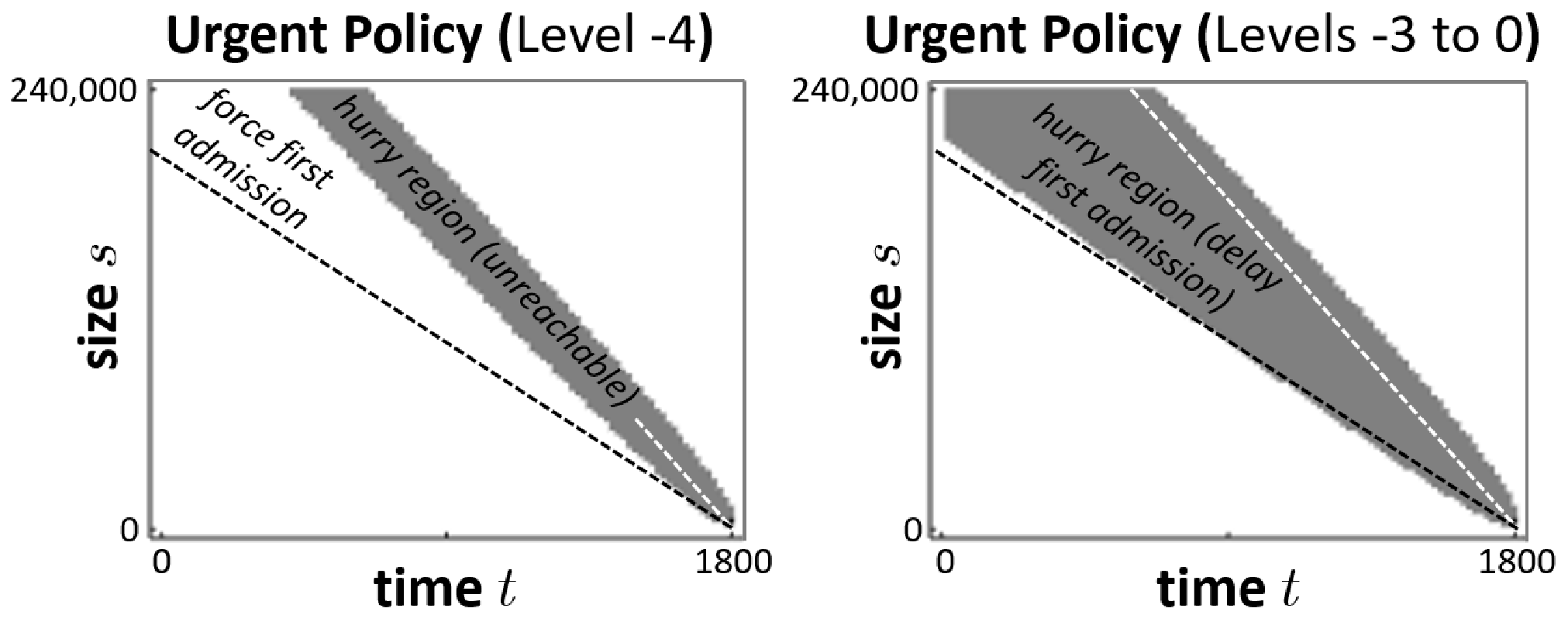} \\
(b)~With Inelastic Requirements of Persistence \& Urgency
\caption{\label{fig:StatefulInelasticResults}  Results of the analysis
of the scenario described in \FigRef{StatelessInelasticResults} upon augmenting
the formulation for richer inelastic objectives.
(a)~All video flows are given
the same simplest persistence requirement, specifically choosing $D_p = 1$ and $\pi_1 = 1$
to strive for no disruption after first admission; while the sample
trajectories no longer exhibit the oscillatory control signals seen in
\FigRef{StatelessInelasticResults}, the video's first admission is
consistently delayed and any need to hurry the elastic flow thereafter
forces the video's permanent suspension. (b)~All video flows
are now also given a simplest urgency requirement, specifically $D_u = 4$ and $\gamma_1 = 1$
so that delay of profitable first admission is at most four stages; while the video's first
admission now always occurs by the fourth stage, its later permanent suspension is also more likely.
The policies in (a) and (b) demonstrate that
persistence and urgency requirements manifest themselves as level-dependent
hurry regions, in our simplest special case bearing strong symmetry across positive and negative
levels, but these details in general are influenced by choice of
the inelastic state dynamics via \FigRef{InelasticStates}.}
\end{figure}

\FigRef{StatefulInelasticResults} presents experiments with only the most basic
persistence and urgency requirements on the video flows. We begin with
persistence alone, so $D_u = 0$, and using the minimal number of levels, so
$D_p = 1$. We take probability $\pi_1 = 1$ so, after first admission, a
transition to permanent suspension occurs immediately upon interruption. Thus,
only patterns without interruption render profitable video in proportion to
its duration in admission. Comparing the sample trajectories in
\FigRef{StatefulInelasticResults}(a) with those of
\FigRef{StatelessInelasticResults}, the achieved persistence is clearly
evident but comes at the expense of delayed first admission and sometimes
also elected permanent suspension as the deadline nears. The behavioral
difference is accomplished by policies with level-dependent hurry regions,
which include the initial state in level~0 but not in level~1. Let us repeat
the experiment but introducing a similarly straightforward urgency requirement:
choose $D_u = 4$ and $\gamma_1 = 1$ so the persistent video is only profitable
if its first admission occurs within 72~seconds (or four stages) of time.
\FigRef{StatefulInelasticResults}(b) shows that, while the video's first
admission now always occurs by the fourth stage, its later permanent suspension
is also more likely. As in the case of persistence, the behavioral difference of
urgency manifests itself as level-dependent hurry regions, which because of the
deterministic increase in per-stage urgency includes the initial elastic state
only at level~-4.

\section{Conclusion \label{sec:Conclu}}
Motivated by increasing demand for multimedia services and the use 
of distributed data centers to globally supply them, a formulation for 
admission control that supports configurable precedence among 
deadline-driven elastic traffic (e.g., scheduled 
synchronization services) and profitable inelastic traffic (e.g., real-time 
streaming services) has been presented. Our formulation can be contrasted 
in several ways from the prevailing literature cited in \SecRef{intro}. Firstly, our scope is only a 
single resource (the link bandwidth) designated primarily 
for elastic traffic (with given size and deadline) but where control affords an 
opportunity to admit profitable inelastic traffic (with given load) if better-than-expected 
network conditions materialize during execution. Secondly, our utility functions 
are time based rather than rate based and, in turn, the controllers can more explicitly 
account for the risk dynamics, which becomes especially advantageous in scenarios with 
long elastic deadlines. The different scope and focus also renders
different algorithmic considerations: the non-convexity of rate-based utility 
functions to represent multi-class traffic and deadlines is a key challenge in the 
prevailing literature, while the key challenge in our formulation is the 
problem discretization so that the employed stochastic shortest path algorithms 
remains tractable. Within its scope, however, our approach features an intriguing
versatility as was demonstrated by the presented experiments e.g., tunable multi-class objectives, 
soft elastic deadlines and congestion avoidance, tunable tradeoff 
between performance for an assumed model and robustness to erroneous models, 
richer inelastic objectives such as persistence or urgency in their admission patterns.
 
An important item for future work is to validate, in simulation or emulation, the extent 
to which the desirable behaviors predicted by our model-based experiments persist in real-world
networks. On the theoretical end, the extent to which our formulation is applicable for scenarios with 
multiple deadline-driven elastic flows is an open question. As 
presented it will not scale beyond a few such flows, but many domains that rely upon 
Markov Decision Processes for large-scale problems report practical successes with  
approximate dynamic programming methods. Future work might also reconsider
the problem within a continuous-state, finite-horizon representation as was highlighted
at the end of \SecRef{ProbForm}. Nearly all of our experiments exhibited optimized policies 
characterized by a small set of stage-dependent thresholds (on the elastic flow's 
remaining size). Perhaps a so-called fluid model, or the limit of our
discrete formulation as the model structure becomes arbitrarily fine-grained, is amenable 
to theoretical analyses in some cases. It is unlikely that
closed-form solutions will exist for the same versatility we examined experimentally, but
proofs under certain special cases may expose structure that promotes greater scalability 
than the discretized formulation presented here. 

%%%%%%%%%%%%%%%%%%%%%%%%%%%%%%%%%%%%%%%%%%%%%%%%%%%%%%%%%%%%%%%%%%%%%%%%%%%%%%%%
%\section*{APPENDIX}

%\newpage
%\bibliography{../bibtex/TextBooks,../bibtex/AdmissionControl}
\bibliography{../bibtex/TextBooks,../bibtex/AdmissionControl}

\end{document}